\documentclass[pra, aps,showpacs,twocolumn,floatfix,superscriptaddress,amsmath]{revtex4-1}
\usepackage{graphicx}
\usepackage{graphics}
\usepackage{braket}
\usepackage{mathtools}
\usepackage[titletoc]{appendix}
\usepackage{appendix}
\usepackage{float}
\DeclarePairedDelimiter\abs{\lvert}{\rvert}%
\makeatletter
\let\oldabs\abs
\def\abs{\@ifstar{\oldabs}{\oldabs*}}
\makeatother
\begin{document}
\title{Noise correlations of two-dimensional Bose gases}
\author{V. P. Singh}
\affiliation{Zentrum f\"ur Optische Quantentechnologien and Institut f\"ur Laserphysik, Universit\"at Hamburg, 22761 Hamburg, Germany}
\affiliation{The Hamburg Centre for Ultrafast Imaging, Luruper Chaussee 149, Hamburg 22761, Germany}
\author{L. Mathey}
\affiliation{Zentrum f\"ur Optische Quantentechnologien and Institut f\"ur Laserphysik, Universit\"at Hamburg, 22761 Hamburg, Germany}
\affiliation{The Hamburg Centre for Ultrafast Imaging, Luruper Chaussee 149, Hamburg 22761, Germany}
\date{\today}
%
%
\begin{abstract}
We analyze density-density correlations of expanding clouds of weakly interacting two-dimensional Bose gases below and above the Berezinskii-Kosterlitz-Thouless transition, with particular focus on short-time expansions. During time-of-flight expansion, phase fluctuations of the trapped system translate into density fluctuations, in addition to the density fluctuations that exist in in-situ. We calculate the correlations of these fluctuations both in real space and in momentum space, and derive analytic expressions in momentum space. Below the transition, the correlation functions show an oscillatory behavior,  controlled by the scaling exponent of the quasi-condensed phase, due to constructive interference. We argue that this can be used to extract the scaling exponent of the quasi-condensate experimentally.
 Above the transition, the interference is rapidly suppressed when the atoms travel an average distance beyond the correlation length. This can be used to distinguish the two phases qualitatively.
\end{abstract}

\maketitle
%
%

\section{Introduction}
Phase coherence is a defining feature of degenerate Bose gases. In three-dimensions (3D), a degenerate Bose gas shows long-range phase coherence, i.e., the single-particle correlation function approaches a constant at large distances. 
 A weakly interacting Bose gas in two-dimensions (2D), however, shows quasi-long-range coherence \cite{mermin_1966, hohenberg_1967}, i.e., the single-particle correlation function decays algebraically.
  As the temperature is increased, the magnitude of the algebraic scaling exponent increases, until it reaches a universal value, at which the system undergoes a phase transition. Above the transition, the correlation function decays exponentially. This transition was predicted by Berezinskii \cite{berezinskii_1972} and by Kosterlitz and Thouless \cite{kosterlitz_1973}, and is known as the Berezinskii-Kosterlitz-Thouless (BKT) transition. It has been observed in several experiments, such as $^4$He films \cite{bishop_1978} and trapped Bose gases \cite{hadzibabic_2006, krueger_2007, clade_2009, choi_2012_2}.

\begin{figure*}[]
\includegraphics[width=0.95\linewidth]{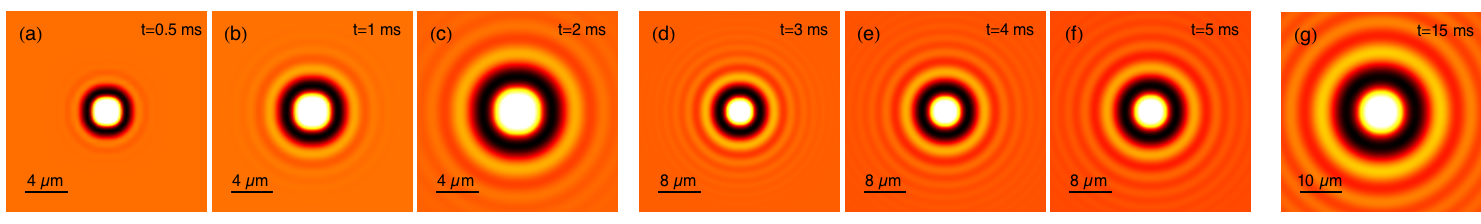} \\
\includegraphics[width=0.95\linewidth]{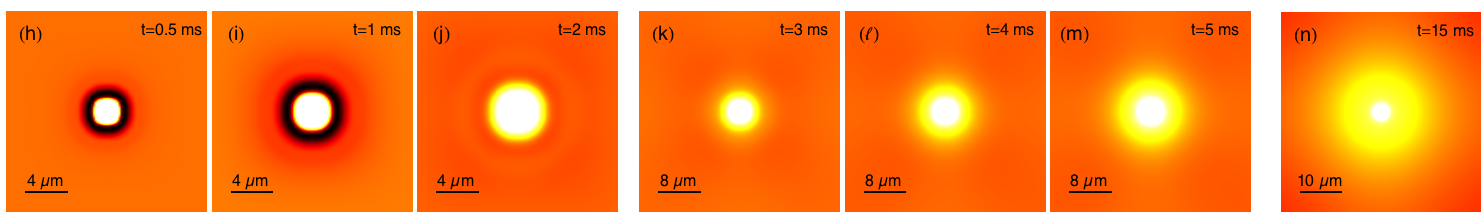}
\caption{(Color online). Evolution of the two-point density correlation function $g_2({\bf r},t)$ is shown after successive expansion times $t$ for the algebraic and exponential regime of a 2D Bose gas of $^{87}\mathrm{Rb}$ atoms. Panels (a)--(g) are for algebraic decay with   scaling exponent $\tau=0.2$ and the short distance cutoff $a= 1\, \mu{\rm m}$. Panels (h)--(n) are for exponential decay with the correlation length $r_0= 1\, \mu{\rm m}$ and $c_0=1$,  see Eq. \ref{g1_hightemp}.}
\label{Fig1}
\end{figure*}

Given the universal importance of correlations, detecting both single-particle and beyond single-particle correlations is of great interest. 
 A seminal study was the measurement of 
 correlations in far-field intensity fluctuations of thermal light sources, observed first in Ref. \cite{brown_1956}, the Hanbury-Brown Twiss effect \cite{fano_1961, glauber_1963}. In analogy to light sources, correlations in density fluctuations have been observed in expanding thermal Bose gases \cite{yasuda_1996, foelling_2005, schellekens_2005, jeltes_2007, hodgman_2011, perrin_2012} and Fermi gases \cite{greiner_2005, rom_2006}. 
The idea of probing many-body states of  ultracold atoms using noise correlations was first suggested in Ref. \cite{altman_2004}. Since then, noise correlations have been used to study quantum phases, such as the Mott-insulator transition in 2D Bose gas \cite{spielman_2007}. 
 Theoretical investigations of noise correlations in 1D Fermi gases were reported in Ref. \cite{mathey_2008, mathey_2009}, of 1D Bose gases in Ref. \cite{imambekov_2009, mathey_2009}, and of 2D Bose gases in Ref. \cite{imambekov_2009, mazets_2012}. The spectrum of these correlations in an expanding 2D Bose gas has been studied experimentally in Ref. \cite{choi_2012, seo_2014}.

In this paper we investigate if and how the BKT transition in ultra-cold atom systems can be detected via noise correlations, and what its signature is.  
 For this purpose, we study the density-density correlations in expanding clouds  of 2D Bose gases, with particular focus on short-time expansions, both above and below the transition temperature. 
 In experiments, the atoms are initially trapped in a trap, and then released by turning off the trap.  
 During the subsequent free ballistic expansion, phase fluctuations present in the trapped system translate into density fluctuations \cite{dettmer_2001}. 
We calculate the correlations of these fluctuations for a homogeneous system  of weakly-interacting bosons in the thermodynamic limit, both in real space and in momentum space, for both the temperature regime of algebraic scaling and of exponential scaling. 
 We find that in the quasi-condensed phase a long-lived interference pattern is visible in the noise correlations, whereas in the thermal phase constructive interference is suppressed after short times of flight. Furthermore, we argue that the shape of the interference pattern and its dependence on the algebraic scaling exponent can be used to determine this exponent experimentally.


This paper is organized as follows: 
In Sec. \ref{detectionscheme} we discuss how the in-situ correlations of the system are related to the density-density correlations of two-dimensional Bose gases in a time-of-flight expansion. 
We first consider only in-situ phase fluctuations, and calculate the density-density correlation function of an expanded cloud of $^{87}$Rb atoms below and above the BKT transition in Sec. \ref{densitycorrelation_section}.
 In Sec. \ref{densitycorrelationspectrum_section} we derive an analytic expression for the spectrum of the density-density correlations of the 2D quasi-condensate (below the BKT transition). We show a comparison to numerical results, and discuss the properties of the analytic solution. 
In Sec. \ref{spectralpeaks_subsection} we analyze the scaling behavior of the peaks for the spectrum of density-density correlations of the condensed phase, and derive an analytic expression for the spectral peak locations. 
In Sec. \ref{spectrumexponential_subsection} we show the spectrum of density-density correlations for the exponential regime of 2D Bose gases (above the BKT transition).
In Sec. \ref{density_fluctuation_section} we expand the analysis to include in-situ density fluctuations of the 2D Bose gas. We analyze their impact on the density-density correlations and the spectrum of these correlations in time-of-flight expansion.
We summarize our results and conclude in Sec. \ref{conclusion_section}.     

%


\section{Correlations in time-of-flight} \label{detectionscheme}
In this section, we relate the density-density correlations in time-of-flight to the in-situ correlations.
 The expansion of the atoms is assumed to be ballistic after release from a tight transverse confinement. Due to the fast expansion in the transverse direction, interaction effects are suppressed quickly during time-of-flight. We consider correlations of the column  density, i.e., averaged over the transverse direction ($z$-axis), which reduces the analysis to two-dimensional expansion. 
 The time evolution of the bosonic field operator is given by \cite{feynman_1965}
\begin{equation}
\hat{\Psi}({\bf r},t) = \int d^2{\bf r}^{\prime} G_2({\bf r}-{\bf r}^\prime, t)\hat{\Psi}({\bf r}^\prime,0).
\end{equation} 
Here, $\hat{\Psi}({\bf r},t)$ is the bosonic single-particle annihilation operator at time $t$, and $\hat{\Psi}({\bf r}^\prime,0)$ is the initial single particle operator. The Green's function of free propagation for a 2D system is defined as
\begin{equation}\label{greenfunc2d}
G_2({\bf r}-{\bf r}^\prime, t) = G_1(x-x^\prime, t) G_1(y-y^\prime, t),
\end{equation}
where
\begin{equation}\label{greenfunc}
G_1(\xi, t)= \sqrt{\frac{m}{2\pi i\hbar t} } \exp \Bigl(i \frac{m \xi^2}{2\hbar t} \Bigr)
\end{equation} 
with $\hbar$ being the reduced Planck constant, $m$ the atomic mass, and $t$ the expansion time. 
We now introduce the two-particle density matrix for the bosonic fields as
\begin{equation} \label{two_particle_matrix}
\rho({\bf r}_1,{\bf r}_2,{\bf r}_3,{\bf r}_4;t)= \braket{ \hat{\psi}^{\dagger}({\bf r}_1,t) \hat{\psi}^{\dagger}({\bf r}_2,t) \hat{\psi}({\bf r}_3,t) \hat{\psi}({\bf r}_4,t) }.
\end{equation} 
 The density-density correlations can be written in terms of the two-particle density matrix as
\begin{equation}\label{correlation_density_relation}
\frac{ \braket{\hat{n}({\bf r}_1,t)\hat{n}({\bf r}_2,t)} }{n({\bf r}_1,t) n({\bf r}_2,t) } = \frac{ \rho({\bf r}_1,{\bf r}_2,{\bf r}_1,{\bf r}_2;t) }{n({\bf r}_1,t) n({\bf r}_2,t) } + \frac{\delta({\bf r}_1-{\bf r}_2)}{n({\bf r}_1,t)},
\end{equation} 
where $n({\bf r},t)=\braket{\hat{n}({\bf r},t)}= \braket{ \hat{\psi}^\dagger({\bf r},t)\hat{\psi}({\bf r},t)}$ is the average density at time $t$. The above expression has two terms on the right-hand side. The first term is the two-point density correlation function, and the second one is the shot-noise contribution that comes from the normal ordering of the bosonic field operators. The two-point density correlation function is related to the two-particle density matrix as $g_2({\bf r}_1,{\bf r}_2;t) \equiv  
\rho({\bf r}_1,{\bf r}_2,{\bf r}_1,{\bf r}_2;t)/(n({\bf r}_1,t) n({\bf r}_2,t))$. For homogeneous systems, $g_2({\bf r}_1,{\bf r}_2;t)$ depends only on the absolute value of the relative distance $|{\bf r}_1-{\bf r}_2|$.
 Therefore, the density-density correlations for a homogeneous 2D system is 
\begin{equation}\label{correlation_density_relation_s}
\frac{ \braket{\hat{n}({\bf r},t)\hat{n}(0,t)} }{n_0^2} = g_2({\bf r},t) + \frac{\delta({\bf r})}{n_0}.
\end{equation}
Here, $n_0$ is the average density for a homogeneous 2D system.
  Next, we calculate the two-point density correlation function $g_2({\bf r}, t)$ of the expanded cloud. The observable measured experimentally is the density-density correlation function, which differs by the shot-noise term. The free evolution of the two-point density correlation function for a 2D system is given by
\begin{align}
g_2({\bf r}_1,{\bf r}_2;t) &= \frac{1}{n_0^2}\int d^2{\bf r}_3 \int d^2{\bf r}^{\prime}_3 \int d^2{\bf r}_4 \int d^2{\bf r}^{\prime}_4  \nonumber \\
& \quad \times  G_2({\bf r}_{13}, t) G_2({\bf r}_{24}, t) G^*_2({\bf r}_{13}^\prime, t) \nonumber\\
& \quad \times G^*_2({\bf r}_{24}^\prime, t) \rho({\bf r}^{\prime}_3,{\bf r}^{\prime}_4,{\bf r}_3,{\bf r}_4;0),
\end{align}
where ${\bf r}_{ij}\equiv{\bf r}_i - {\bf r}_j$ and ${\bf r}_{ij}^\prime\equiv{\bf r}_i - {\bf r}_j^\prime$.  
  Using the translational invariance of the two-point density correlation function, and Eqs. (\ref{greenfunc2d}) and (\ref{greenfunc}), we get the following expression: 
\begin{align}
g_2({\bf r}_{12},0;t) &= \frac{1}{n_0^2} \left(\frac{m}{4\pi\hbar t} \right)^2 \int d^2{\bf r} \int d^2{\bf r}^{\prime} \exp \Bigl( i \frac{m}{4\hbar t}  \nonumber \\
& \quad \times  \bigl[ ({\bf r}_{12}-{\bf r})^2 - ({\bf r}_{12}-{\bf r}^\prime)^2 \bigr] \Bigr)  \nonumber\\
& \quad \times  \rho_r({\bf r}^{\prime};{\bf r};0),
\end{align}
where $\rho_r({\bf r}^{\prime};{\bf r};t) \equiv \rho\left(\frac{{\bf r}^\prime}{2},-\frac{{\bf r}^\prime}{2}, \frac{{\bf r}}{2}, -\frac{{\bf r}}{2};t\right)$. In the above expression, we have used the center of mass transformation, hence the reduced mass $m/2$ as compared to the expansion of a single atom in Eq. (\ref{greenfunc}). 
  After  Fourier transforming the Green's functions, we write this as 
\begin{align}\label{correlation_expression_general}
g_2({\bf r},t) &= \frac{1}{(2\pi n_0)^2} \int d^2{\bf q} \int d^2{\bf R} \cos {\bf q\cdot r} \cos {\bf q\cdot R} \nonumber\\
& \times \Bigg\langle \hat{\psi}^{\dagger}\left(\frac{\hbar{\bf q}t}{m},0 \right) \hat{\psi}^{\dagger}({\bf R},0)  \hat{\psi}\left({\bf R}+ \frac{\hbar{\bf q}t}{m},0 \right) \hat{\psi}({\bf 0},0)\Bigg\rangle,
\end{align}
which gives the two-point density correlation function at time $t$.

\begin{figure}[]
\hspace{-2.5 in}(a) \\
 \includegraphics[width=0.80\linewidth]{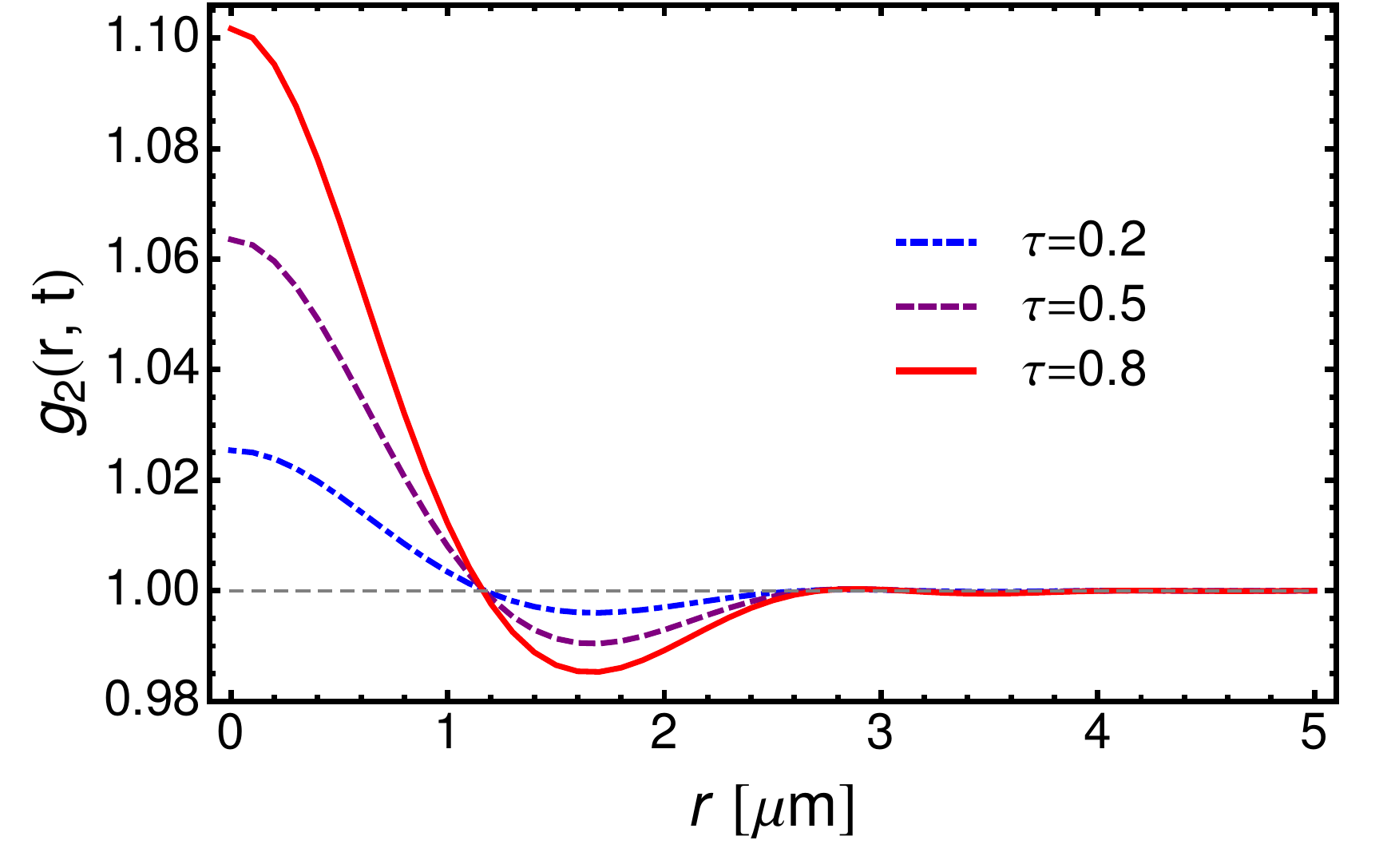} \\
\hspace{-2.5 in}(b) \\
 \includegraphics[width=0.80\linewidth]{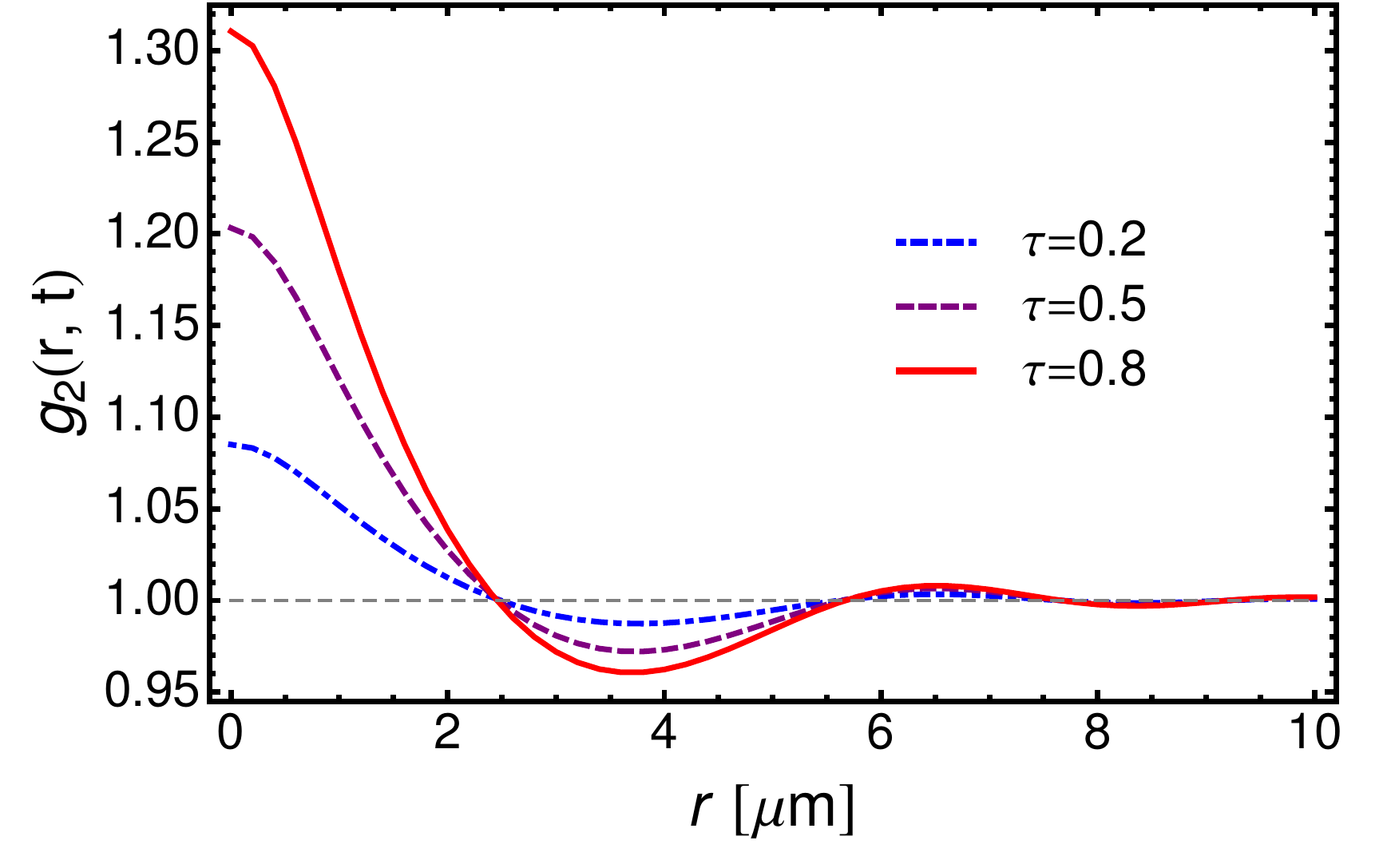}
\caption{(Color online). The density correlation function $g_2({\bf r},t)$ of an expanding cloud of $^{87}\mathrm{Rb}$ atoms below the BKT transition (quasi-condensed phase) after two different expansion times $t$, 
(a) $t=0.5\, {\rm ms}$, and
(b) $t=3\, {\rm ms}$.
The blue, dot-dashed line corresponds to the scaling exponent $\tau=0.2$; the purple, dashed line corresponds to $\tau=0.5$; the red, solid line corresponds to $\tau=0.8$.}
\label{Fig2}
\end{figure}
\begin{figure}[]
\hspace{-2.5 in}(a) \\
 \includegraphics[width=0.8\linewidth]{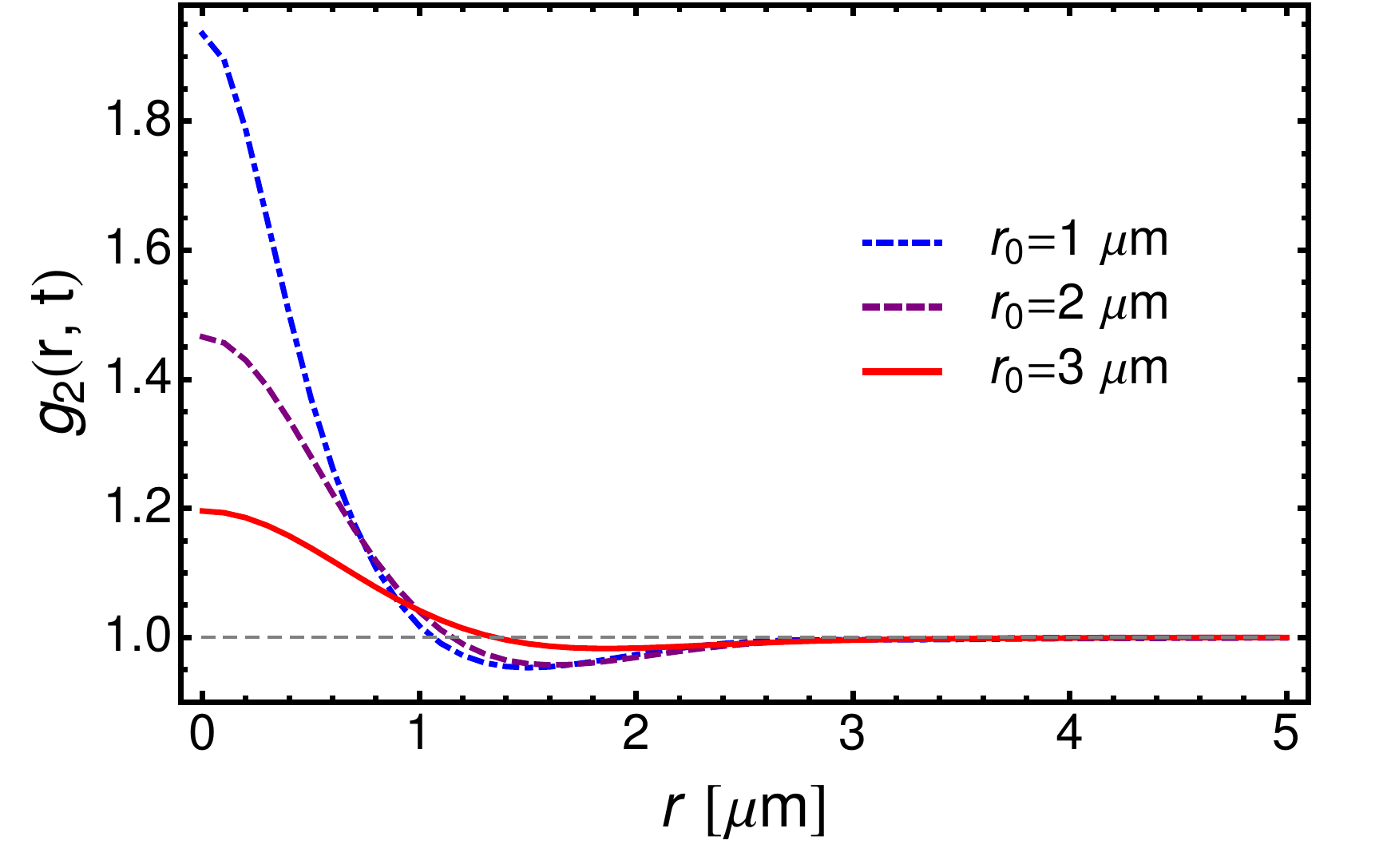} \\
\hspace{-2.5 in}(b) \\
 \includegraphics[width=0.8\linewidth]{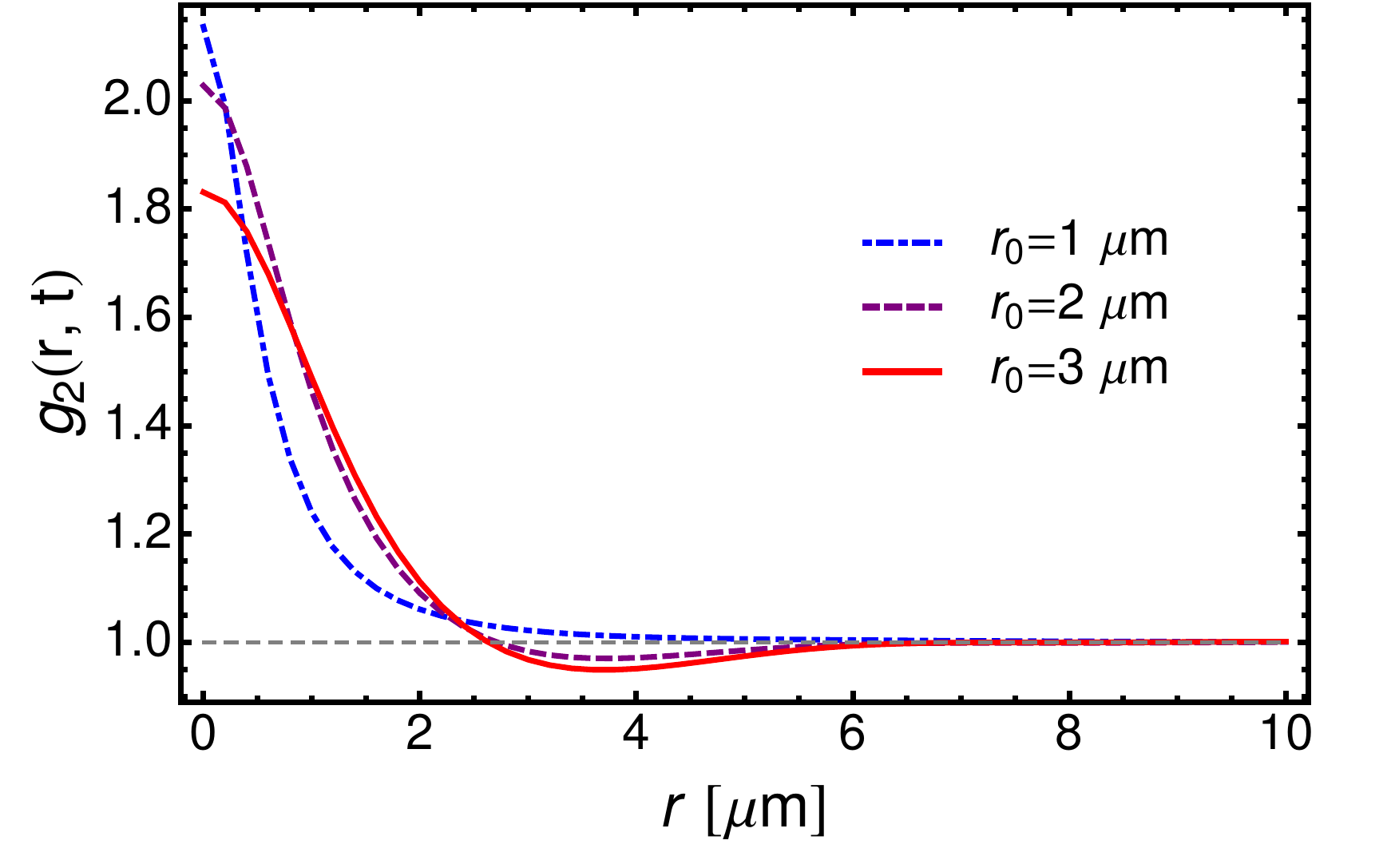}
\caption{(Color online). The correlation function $g_2({\bf r},t)$ of an expanded cloud of $^{87}\mathrm{Rb}$ atoms above the BKT transition (thermal phase) after two different expansion times $t$, 
(a) $t=0.5\, {\rm ms}$, and
(b) $t=3\, {\rm ms}$.
The blue, dot-dashed line corresponds to the correlation length $r_0=1\,\mu {\rm m}$; the purple, dashed line corresponds to  $r_0=2\,\mu {\rm m}$; the red, solid line corresponds to  $r_0=3\,\mu {\rm m}$.
We set $c_0$ to be $1$. }
\label{Fig3}
\end{figure}
%

%
%
%
\subsection*{2D Bose gases} 
A two-dimensional Bose gas in equilibrium undergoes the BKT phase transition, defined through the long-range behavior of the two-point correlation function. The two-point correlation function of the field is defined as
\begin{equation}
g_1({\bf r}) = \frac{\braket{\hat{\psi}^{\dagger}({\bf r}) \hat{\psi} ({\bf 0})} }{n_0}. 
\end{equation}
As discussed in Ref. \cite{mora_2003} we write the single particle operator as $\Psi({\bf r}) \approx \exp(i \hat{\theta}({\bf r})) \sqrt{\hat{n}({\bf r})}$.
 $\hat{n}({\bf r})$ is the density operator and $\hat{\theta}({\bf r})$ is the phase. 
  We write the density operator as $\hat{n}({\bf r})= n_0 +\delta\hat{n}({\bf r})$, where $\delta\hat{n}({\bf r})$ is the operator of density fluctuations and $n_{0}$ is the average density. For now, we neglect density fluctuations and only consider the  phase fluctuations of the gas. 
  We will investigate the corrections due to  density fluctuations in Sect. \ref{density_fluctuation_section}. 
   Thus, the single particle operator is approximated by $\Psi({\bf r}) \approx  \sqrt{n_{0}}  \exp(i \hat{\theta}({\bf r}))$.
    As described in Ref. \cite{mora_2003} the single particle correlation function
     is approximately $g_1({\bf r}) = n_{0} \exp(- \langle \Delta \theta({\bf r})^{2}\rangle/2)$, where $\Delta \theta({\bf r}) \equiv \theta({\bf r}) - \theta({\bf 0})$. 
         In the quasi-condensed phase, the correlation function of the phase scales logarithmically. Therefore,
          the two-point correlation function decays algebraically as  
\begin{equation} \label{g1_quasicondensed}
g_1({\bf r}) \approx \left(\frac{a^2}{a^2 + |{\bf r}|^2}\right)^{\tau/8} \equiv \mathcal{F}_{a}({\bf r}) ,
\end{equation}
where $a$ is a short distance cutoff \cite{distancecutoff_a}. $\tau$ is the scaling exponent \cite{prokofev_2001,prokofev_2002}, which varies from $0$ to $1$ as the temperature is increased from $0$ to the Kosterlitz-Thouless temperature $T_{c}$. 
Above the critical temperature, the correlation function of the phase scales linearly, and therefore the functional form of the two-point correlation changes to exponential decay as
\begin{equation} \label{g1_hightemp}
g_1({\bf r}) \approx \Biggl( \frac{c_0^2}{c_0^2 + 4 \sinh^2(|{\bf r}|/r_0)} \Biggr)^{1/2} \equiv \mathcal{F}_e({\bf r}),
\end{equation}  
where $r_0$ is the correlation length.
	To  determine the two-particle density matrix we evaluate 
\begin{eqnarray}\label{densitymatrix_factorization}
\rho ({\bf r}_1, {\bf r}_2, {\bf r}_3, {\bf r}_4)
 & \sim n_0^2 \langle e^{-i \hat{\theta}({\bf r}_1)-i \hat{\theta}({\bf r}_2)+i \hat{\theta}({\bf r}_3)+i \hat{\theta}({\bf r}_4)}\rangle.
\end{eqnarray}
As described in Ref. \cite{mathey_2009} this gives
\begin{eqnarray}\label{densitymatrix_factorization}
\rho ({\bf r}_1, {\bf r}_2, {\bf r}_3, {\bf r}_4)
 &\sim n_0^2 \frac{\mathcal{F}_i({\bf r}_{13}) \mathcal{F}_i({\bf r}_{14}) \mathcal{F}_i({\bf r}_{23}) \mathcal{F}_i({\bf r}_{24}) }{\mathcal{F}_i({\bf r}_{12}) \mathcal{F}_i({\bf r}_{24})},
\end{eqnarray}
where $i$ stands for algebraic or exponential regime as $i =a, e$.
  Therefore, the two-point density correlation function for the phase fluctuating 2D Bose gas is  
\begin{align}\label{correlation_expression}
g_2({\bf r},t) &= \frac{1}{(2\pi)^2} \int d^2{\bf q} \int d^2{\bf R} \cos {\bf q\cdot r} \cos {\bf q\cdot R} \nonumber\\
& \quad \times \Biggl( \frac{\mathcal{F}_i({\bf q}_t)^2 \mathcal{F}_i({\bf R})^2}{\mathcal{F}_i({\bf R} - {\bf q}_t) \mathcal{F}_i({\bf R}+ {\bf q}_t)}  \Biggr),
\end{align}
 where ${\bf q}_t \equiv \hbar {\bf q}t/m$.

%
\section{Real space density correlations} \label{densitycorrelation_section}
  In this section, we discuss  $g_2({\bf r},t)$ of a cloud of $^{87}\mathrm{Rb}$ atoms below and above the BKT transition. We evaluate   $g_2({\bf r},t)$ numerically by using Eq. (\ref{correlation_expression}). For a finite system $L$ we replace the components $x$ and $y$ of ${\bf r} = (x,y)$ by $x \rightarrow L/\pi \sin(\pi x/L)$, and analogously for $y$. We calculate  $g_2({\bf r},t)$ for the quasi-condensed phase (below the BKT transition) with three different scaling exponents $\tau$ after $0.5\, {\rm ms}$ and $3\, {\rm ms}$ expansion times, see Fig. \ref{Fig2}. We set the short distance cutoff $a$ to be $1\, \mu {\rm m}$.
In Fig. \ref{Fig3} we show $g_2({\bf r},t)$  for the thermal phase (above the BKT transition) for three different correlation lengths $r_0$ after $0.5\, {\rm ms}$ and $3\, {\rm ms}$ expansion times .
\begin{figure}[] 
\hspace{-2.5 in}(a) \\
\includegraphics[width=0.80\linewidth]{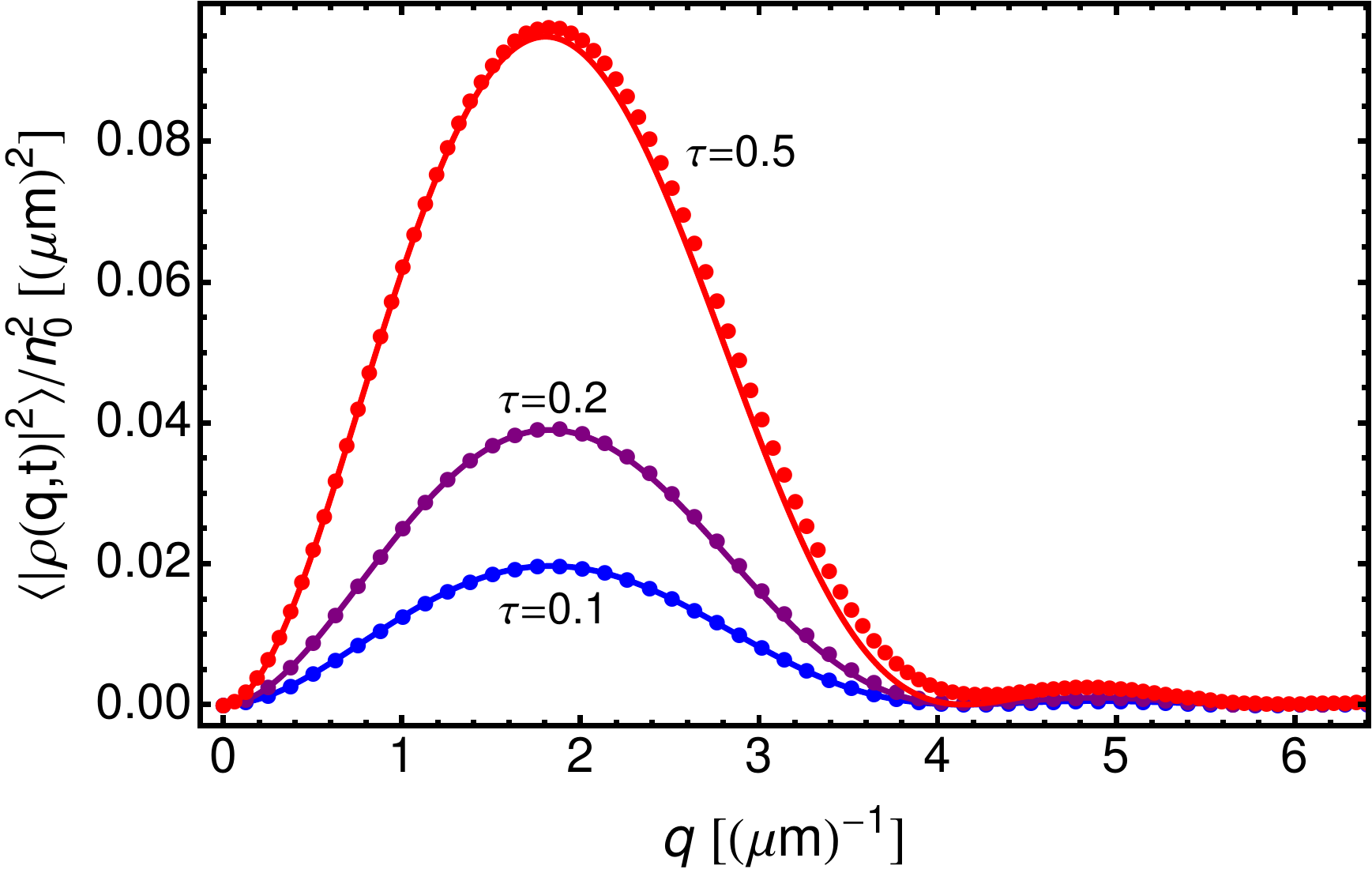} \\
\hspace{-2.5 in}(b)\\
\hspace{0.06 in}  \includegraphics[width=0.80\linewidth]{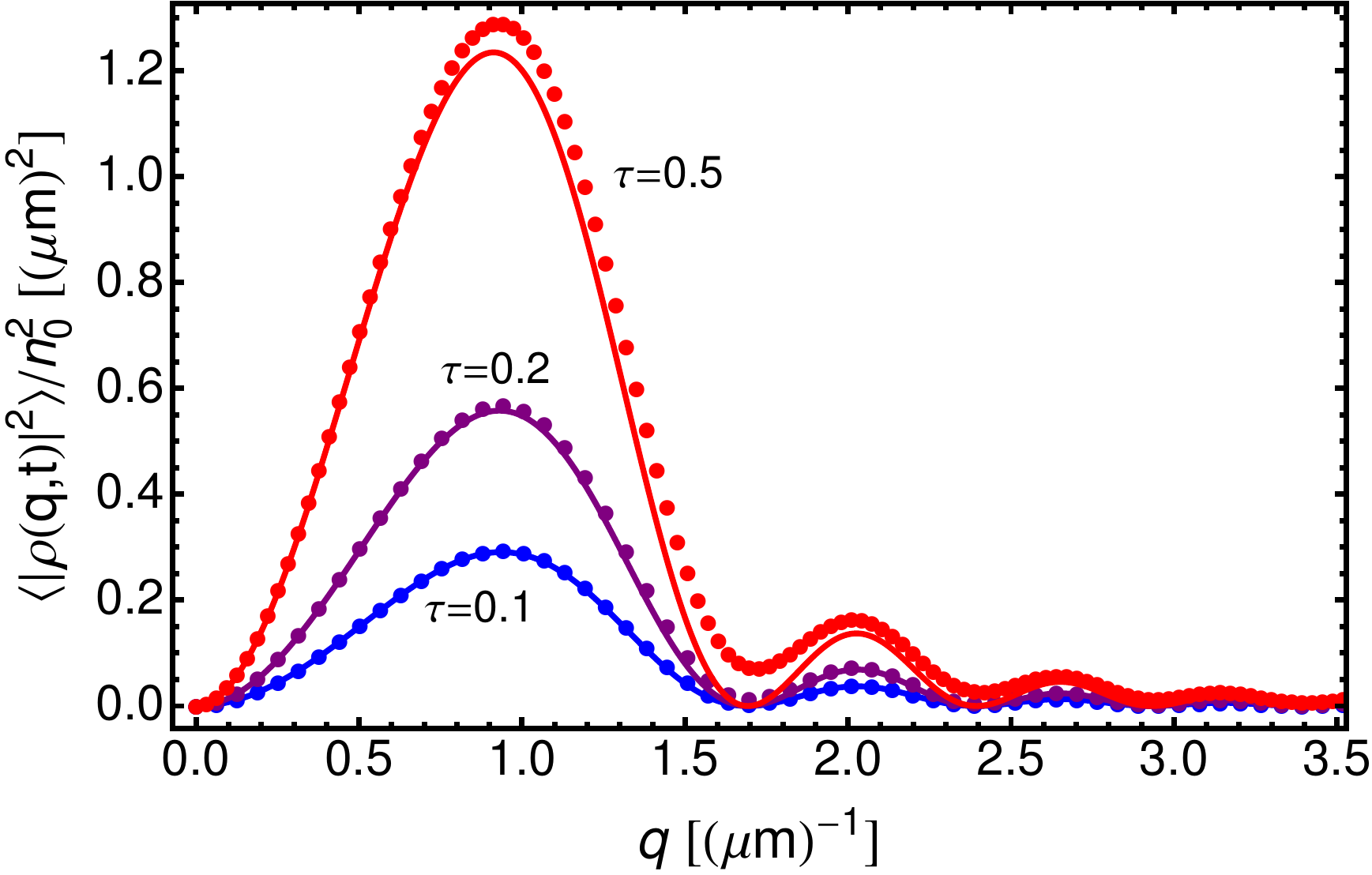}
\caption{(Color online). The spectrum of the density-density correlations $\braket{|\rho({\bf q})|^2}/n_0^2$ of an expanded cloud of $^{87}\mathrm{Rb}$ atoms below the BKT transition (quasi-condensed phase) after two different expansion times $t$.
(a) $t=0.5\, {\rm ms}$, and 
(b) $t=3\, {\rm ms}$.  
The blue, solid line corresponds to $\tau=0.1$; the purple, solid line corresponds to $\tau=0.2$; the red, solid line corresponds to $\tau=0.5$.  
These lines are the analytic results obtained from Eq. (\ref{spectrum_1order}). The filled circles are numerical results obtained for a finite size system from Eq. (\ref{spectrum_expression}).}
\label{Fig4}
\end{figure}

 Comparing  Figs. \ref{Fig2} and \ref{Fig3} we again see an oscillatory behavior due to interference, which is also shown in Fig. \ref{Fig1}. 
   This interference pattern persists for longer expansion times for the quasi-condensed phase, while it vanishes for the thermal phase after a short expansion time. It vanishes earlier, if the the correlation length is small, as can be seen from Fig. \ref{Fig3}.
  Furthermore, we observe that the overall magnitude of $g_2({\bf r},t) -1$ increases with increasing $\tau$ below the BKT transition, and  with decreasing correlation length $r_0$ above the transition. Both cases, increasing $\tau$ and decreasing $r_{0}$, describe increasing temperature, and therefore an overall increase of the noise level.

\begin{figure}[]
\includegraphics[width=0.80\linewidth]{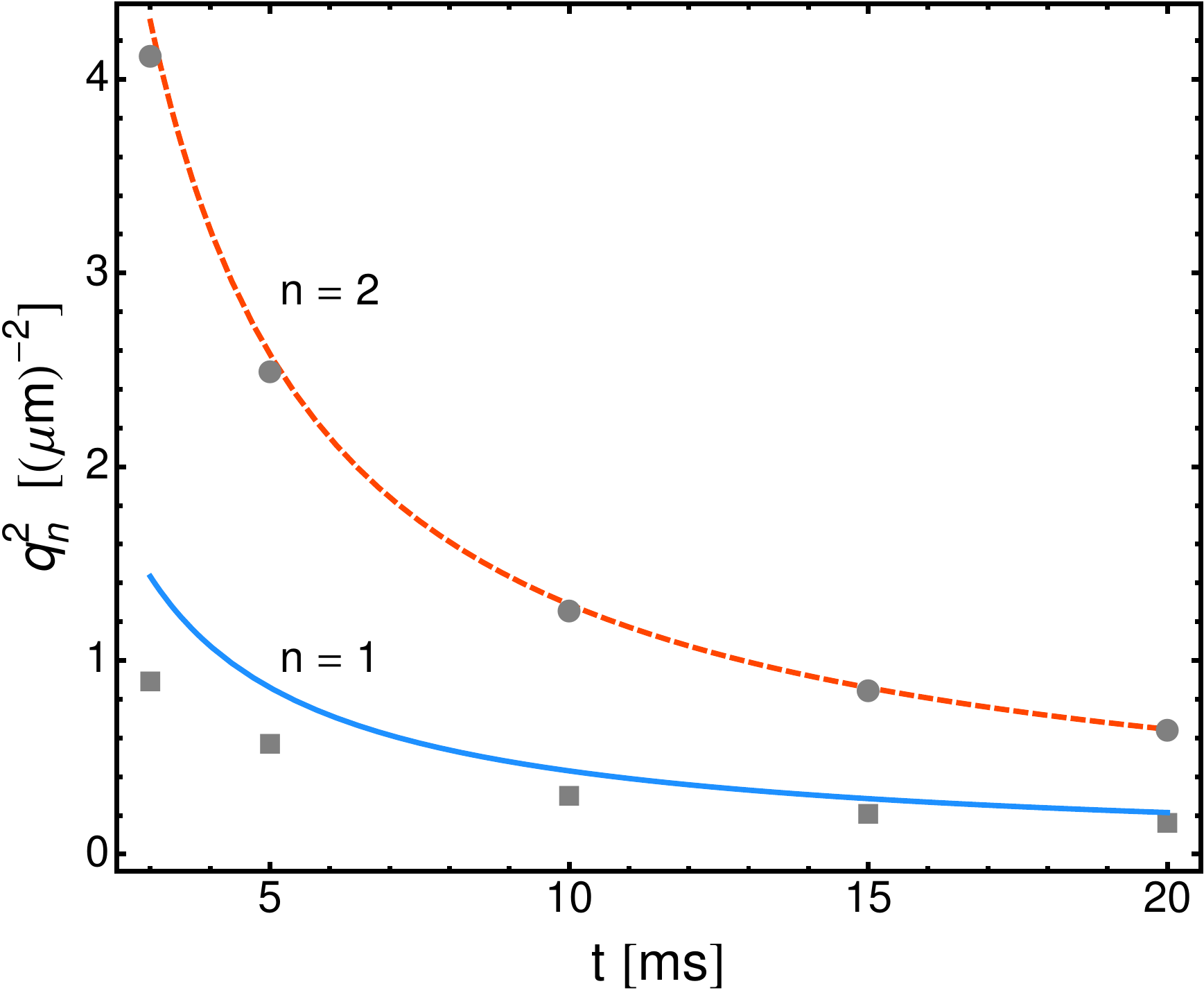}
\caption{(Color online). Comparison of mean-field spectral peak locations to the peak locations obtained numerically from Eq. (\ref{spectrum_expression}) for the quasi-condensed phase with exponent $\tau=0.1$, after several expansion times $t$.
The light blue, solid line and the orange, dashed line are the square of wavevectors corresponding to the first $(n=1)$ and the second $(n=2)$ spectral peak, respectively, obtained by the mean-field results, Eq. (\ref{zero-order_peakpositions}), for $3 \sim 20\, {\rm ms}$ expansion times. The filled squares and the filled circles are the numerical results corresponding to $n=1$ and $n=2$ spectral peak, respectively, obtained numerically from Eq. (\ref{spectrum_def}) for a finite system.}
\label{Fig5}
\end{figure}
\begin{figure}[]
\includegraphics[width=0.90\linewidth]{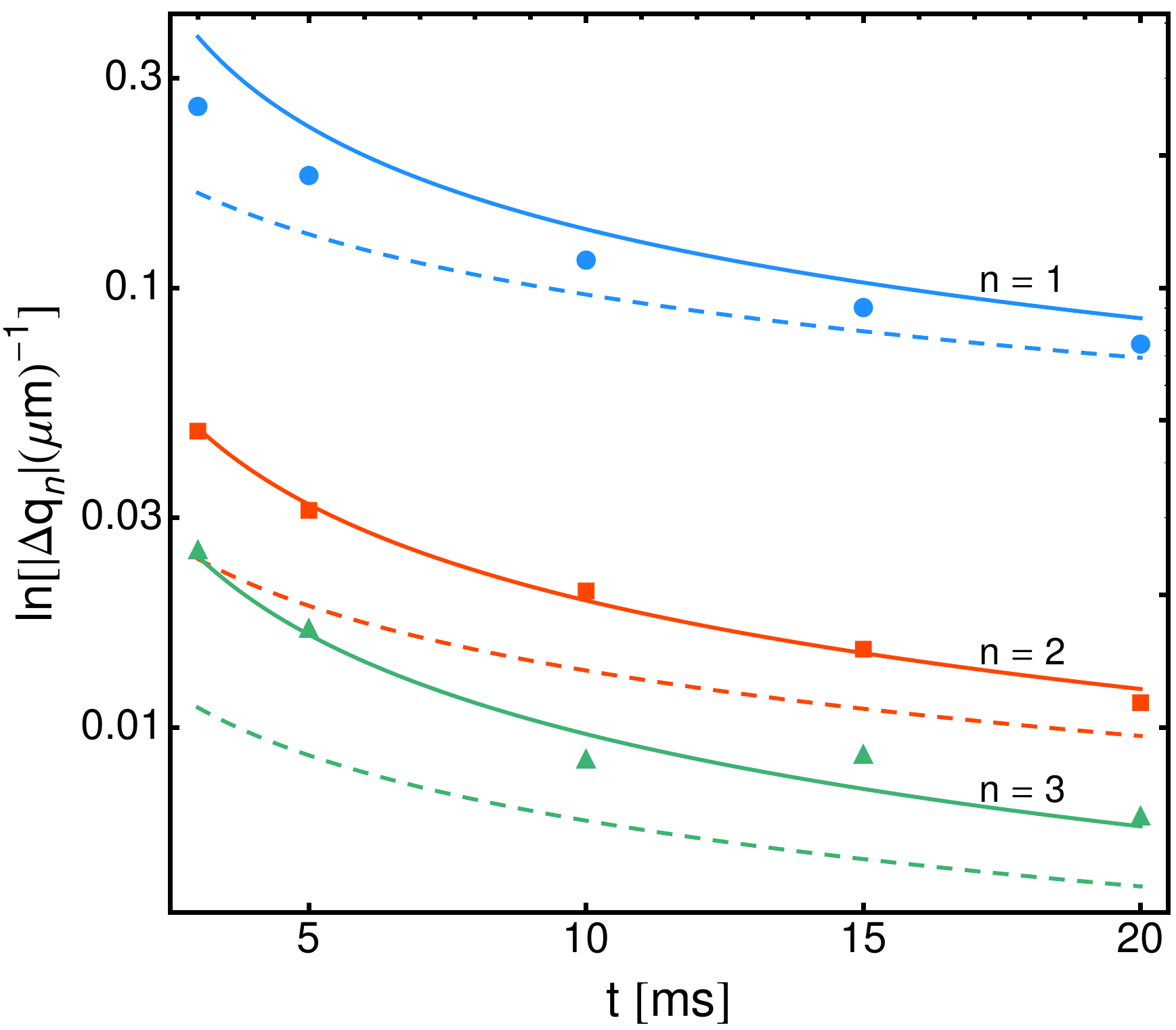}
\caption{(Color online). The shift $\Delta q_n$ of spectral peak locations of the density-density correlations for the quasi-condensed phase with exponent $\tau=0.1$ after various expansion times $t$. The filled circles, the filled squares, and the filled triangles are results for the shift corresponding to the first $(n=1)$, the second $(n=2)$, and the third $(n=3)$ spectral peak, respectively, obtained numerically from Eq. (\ref{spectrum_expression}) for a finite system.   
The solid lines are shifts of the corresponding spectral peaks, obtained from Eq. (\ref{peak_shift}) and plotted as function of expansion time. 
The dashed lines are shifts of the corresponding spectral peaks, obtained from Eq. (\ref{approximated_peak_positions}). }
\label{Fig6}
\end{figure}
%
%
%
\section{Correlations in momentum space} \label{densitycorrelationspectrum_section}
 In this section, we discuss  the power spectrum density-density correlations, defined as \cite{asymptotic_g2realspace}
\begin{equation} \label{spectrum_def}
\frac{\braket{|\rho({\bf q},t)|^2} }{n_0^2} =\int d^2{\bf r}\cos {\bf q\cdot r} [g_2({\bf r},t) - 1],
\end{equation}
which is essentially the Fourier transform of $g_2({\bf r},t)$. We thus Fourier transform Eq. (\ref{correlation_expression}) and get
\begin{align}\label{spectrum_expression}
\frac{\braket{|\rho({\bf q})|^2}}{n_0^2} = \int d^2{\bf r} \cos {\bf q\cdot r} \Biggl( \frac{\mathcal{F}_i({\bf q}_t)^2 \mathcal{F}_i({\bf r})^2}{\mathcal{F}_i({\bf r} - {\bf q}_t) \mathcal{F}_i({\bf r}+ {\bf q}_t)}   -1 \Biggr). 
\end{align}
 This quantity can be obtained experimentally by determining the density-density correlations for a single realization, and then repeating the measurement and average over realizations.
%
%
  We first derive an analytic expression for the spectrum of density-density correlations for the quasi-condensed phase by expanding the two-point correlation function, Eq. (\ref{g1_quasicondensed}), to first order in $\tau$ (see Appendix A for details). We get the following analytic result:
\begin{align}\label{spectrum_1order}
\frac{\braket{|\rho({\bf q})|^2}}{n_0^2} &\approx \frac{\pi a \tau K_1(a q)}{q} \left(\frac{a^2}{a^2+\frac{q^2 \hbar ^2 t^2}{m^2}}\right)^{\tau /4} \nonumber\\
& \quad \times \left(1 - \cos \left(\frac{q^2 \hbar t }{m}\right) \right), 
\end{align}  
where $K_1$ is the Bessel function of second kind and $q\equiv|{\bf q}|$. The above expression is a product of three terms on the right-hand side. The first one is an exponential decay with the short distance cutoff $a$ for the quasi-condensed phase of the Bose gas. This term is time-independent, and purely a result of the system having a short range cutoff.  The third one is the mean-field term, see  \cite{imambekov_2009}. If the system had perfect coherence, this term generates an undamped interference pattern.
 The second term is due to quasi-long-range order: As the atomic cloud expands, the atoms interfere with other atoms from the system further and further apart. As a result the constructive interference diminishes following a power-law.

  We evaluate the spectrum of the density-density correlations by numerical integration of Eq. (\ref{spectrum_expression}) for a finite system and compare these results to the analytic results obtained from Eq. (\ref{spectrum_1order}) for a cloud of $^{87}\mathrm{Rb}$ atoms below the BKT transition. We consider a two-dimensional finite system of length $L$ for the numerical calculations. Figure  \ref{Fig4} shows a direct comparison of the analytic results to the numeric ones for three different temperature dependent exponents $\tau$ after $0.5\, {\rm ms}$ and $3\, {\rm ms}$ expansion times. Both numeric and analytic results are in good agreement, especially for small exponents $\tau$.   

  We observe that the magnitude of the spectral peaks increases with both the expansion time and the scaling exponent $\tau$. 
In particular for small momenta, corresponding to large distances, we can expand Eq. (\ref{spectrum_1order}) and obtain:
\begin{eqnarray}\label{spectrum_1order_smallq}
\frac{\braket{|\rho({\bf q})|^2}}{n_0^2} &\approx& \pi \tau \frac{\hbar^{2} q^{2} t^{2}}{2 m^{2}}.
\end{eqnarray}  
We note that the dependence on $a$ drops out, and that the overall magnitude of the noise just scales linearly in $\tau$. We propose to detect $\tau$ by fitting the power spectrum with either Eq. (\ref{spectrum_1order}) or Eq. (\ref{spectrum_1order_smallq}).
\begin{figure}[]
\hspace{-2.5 in}(a) \\
\includegraphics[width=0.80\linewidth]{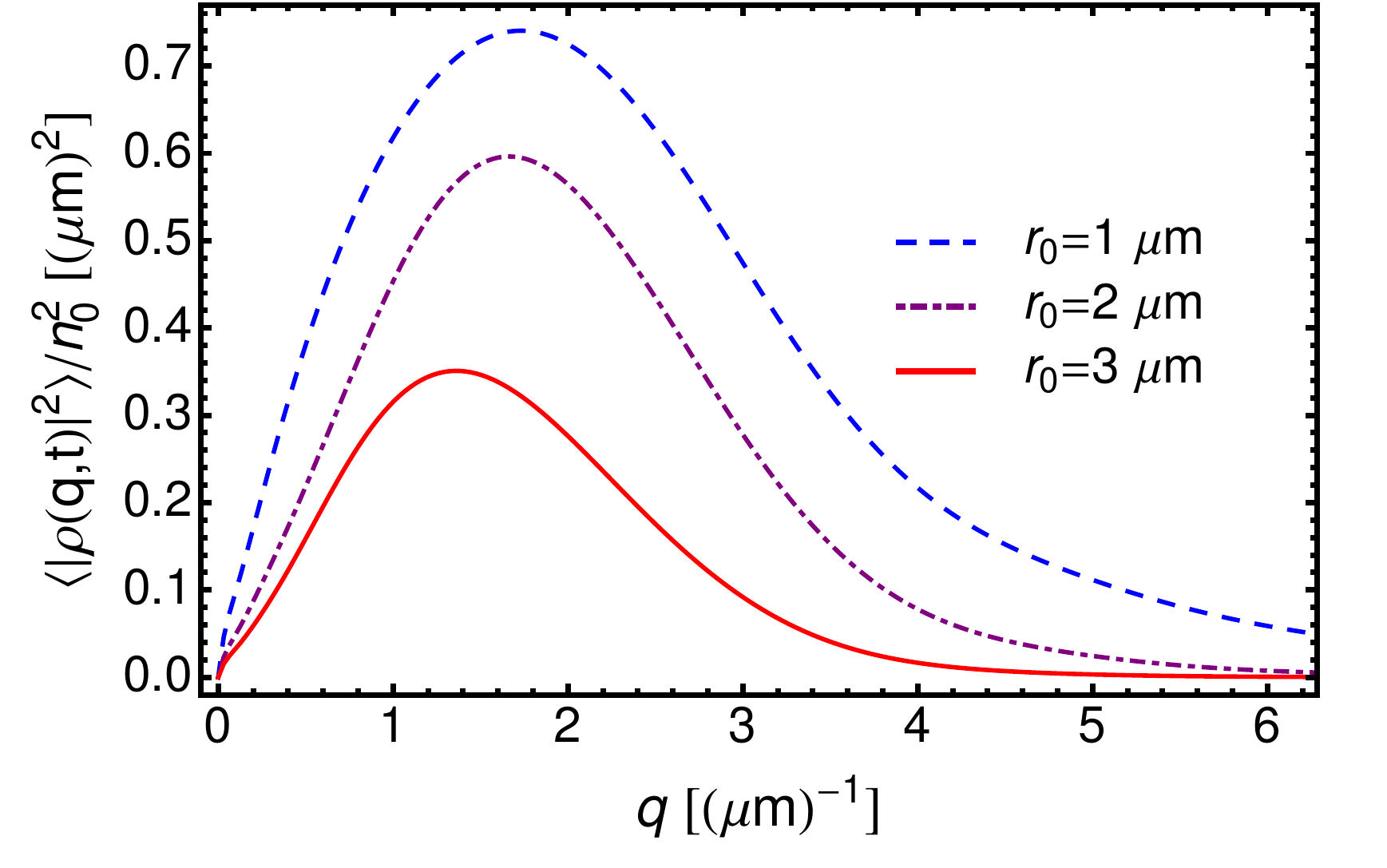} \\
\hspace{-2.5 in}(b)\\
\hspace{0.06 in} \includegraphics[width=0.80\linewidth]{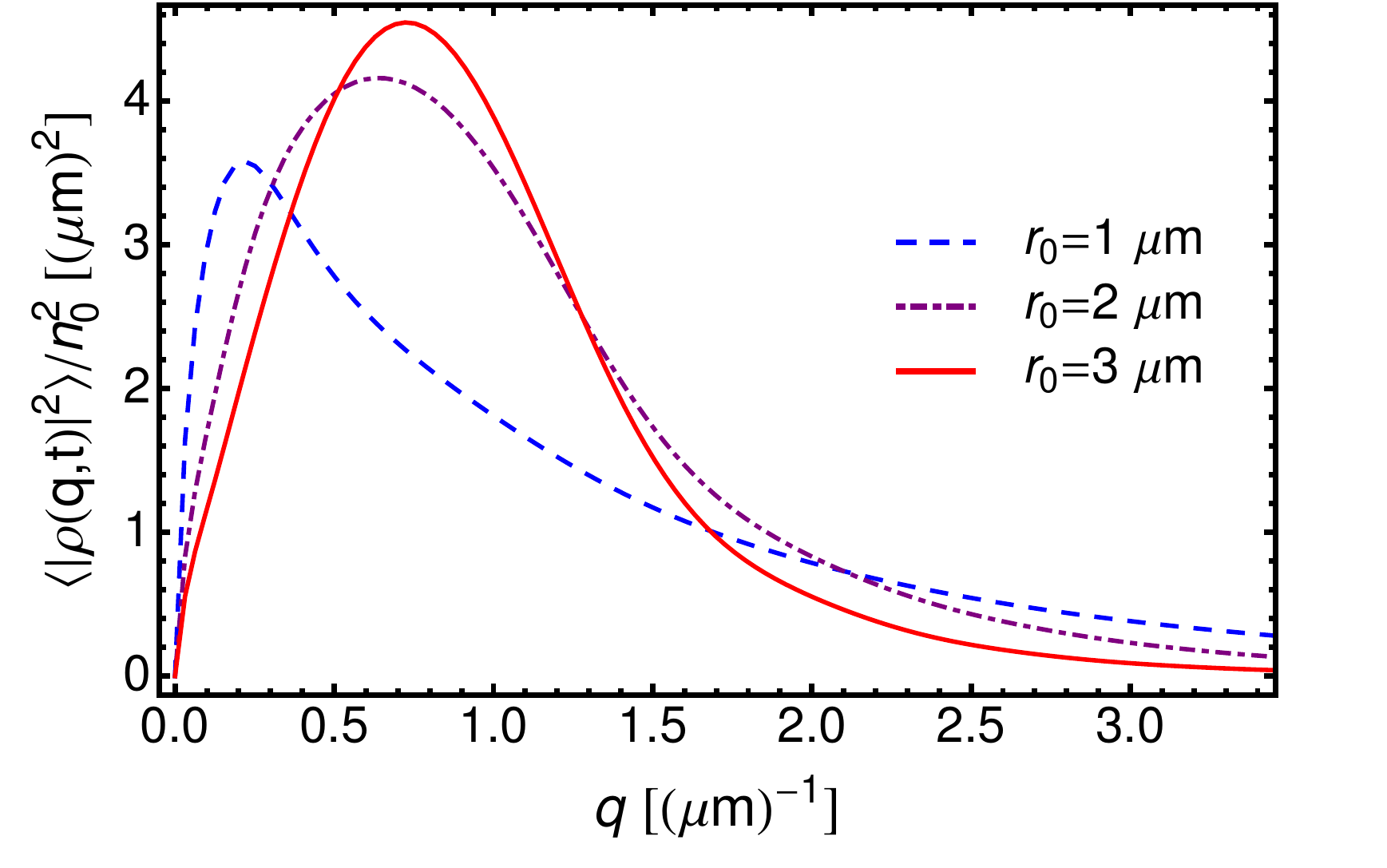}
\caption{(Color online). The spectrum of the density-density correlations $\braket{|\rho({\bf q})|^2}/n_0^2$ for a cloud of $^{87}\mathrm{Rb}$ atoms above the BKT transition (thermal phase) after two different expansion times $t$.
(a) $t=0.5\, {\rm ms}$, and
(b) $t=3\, {\rm ms}$.
The blue, dashed line corresponds to the correlation length $r_0= 1\, \mu {\rm m}$; the purple, dot-dashed line corresponds to $r_0= 2\, \mu {\rm m}$; the red, solid line corresponds to $r_0= 3\, \mu {\rm m}$. We set $c_0$ to be $1$. }
\label{Fig7}
\end{figure}
%
%
%
%
\subsection{Spectral peak locations} \label{spectralpeaks_subsection}
We now discuss the scaling behavior of the spectral peaks, as shown in Fig. \ref{Fig4}, and derive analytic expressions for the spectral peak locations from Eq. (\ref{spectrum_1order}).  
  The shift of the spectral peak locations owing to the temperature and the expansion time is given by (see Appendix B) 
\begin{align} \label{peak_shift}
\Delta q_{n} &\approx  q_{n0} L_n^2 \Bigl(q_{n0} t^2\tau \hbar^2 K_1(aq_{n0}) + 2a L_n^2 m^2 K_2(aq_{n0}) \Bigr) \nonumber\\
& \quad \times \Bigl[ q_{n0} \Bigl(2a^2 m^2(a^4 - q_{nt}^4) + t^2  \bigl(-2a^2 \tau   \nonumber \\
&\quad +L_n^2(\tau- 4q_{n0}^2 q_{nt}^2 ) \bigr) \hbar^2 \Bigr) K_1(aq_{n0}) + 2 a m^2 L_n^2         \nonumber \\
& \quad  \times \bigl(3L_n^2 +q_{nt}^2\tau \bigr) K_2(aq_{n0}) \Bigr]^{-1},
\end{align}
where $K_1$ and $K_2$ are the Bessel functions of the second kind, and $L_n^2(t)= a^2 +q_{nt}^2$, where $q_{nt} \equiv \frac{q_{n0}\hbar t}{m}$. Here, $q_{n0}$ is defined as
\begin{equation} \label{zero-order_peakpositions}
q_{n0} = \sqrt{\frac{(2n-1)\pi m}{\hbar t}},
\end{equation}
which is the location for the spectral peaks obtained from the mean-field term in Eq. (\ref{spectrum_1order}). $n$ is the peak order number.

  To analyze Eq. (\ref{peak_shift}), we consider two different cases for the shift of the spectral peak locations as either $a q_{n0} \ll 1$ or $a q_{n0} \gg 1$. For $a q_{n0} \ll 1$, i.e., $t \gg \frac{m a^2}{\hbar} (2n-1)\pi$, a simplified expression for the spectral peak shift is given by (see Appendix B for details)
\begin{equation}\label{approximated_peak_positions}
\Delta q_{n} \sim  - \frac{q_{n0}(4 + \tau)}{ 4(2n-1)^2 \pi^2 + \frac{2ma^2}{\hbar t}(2n-1)\pi -5\tau -12}, 
\end{equation}
which is the first-order correction to the spectral peak locations.

\begin{figure}[]
\hspace{-2.5 in}(a) \\
 \includegraphics[width=0.80\linewidth]{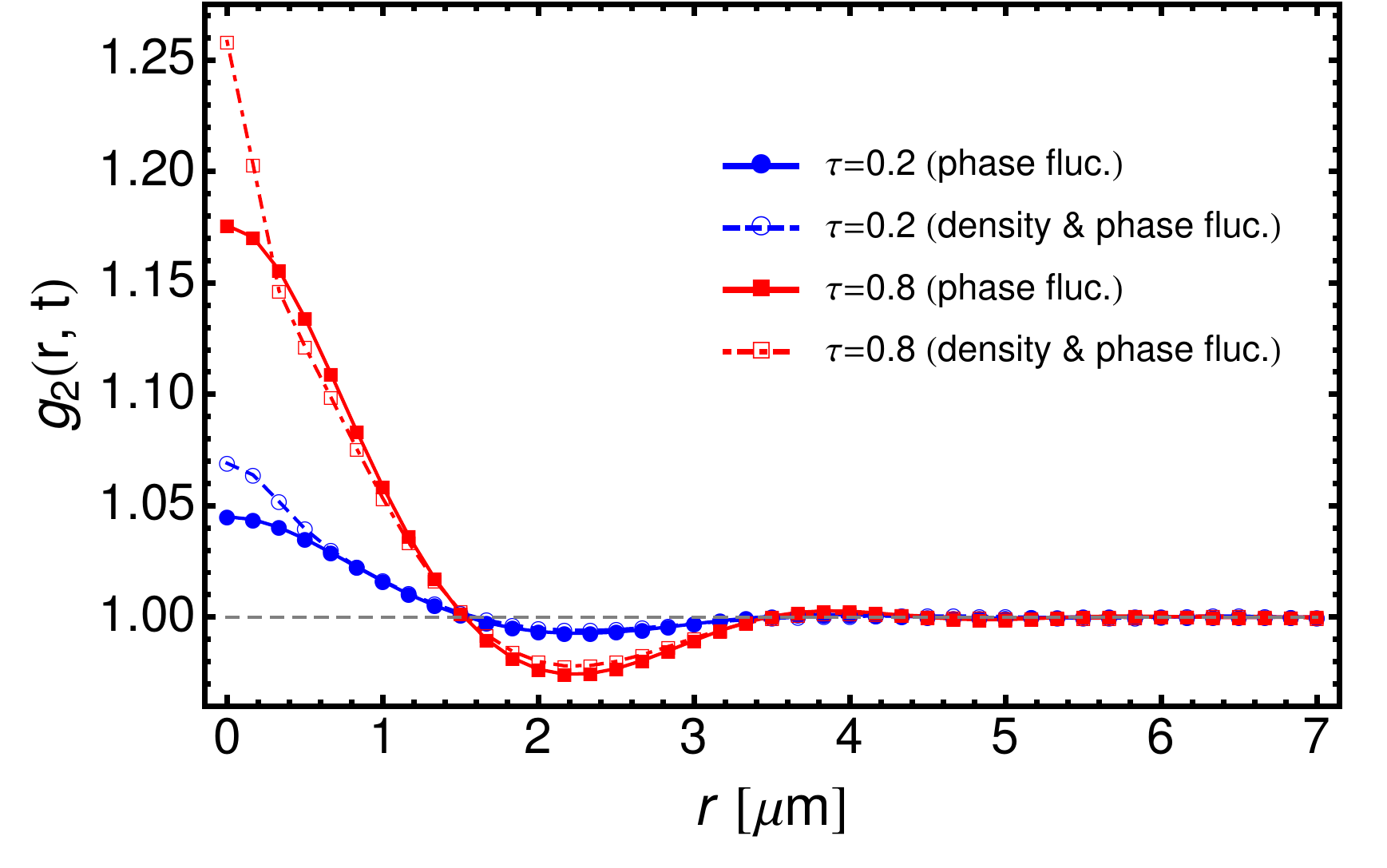} \\
\hspace{-2.5 in}(b) \\
\hspace{0.06 in}  \includegraphics[width=0.80\linewidth]{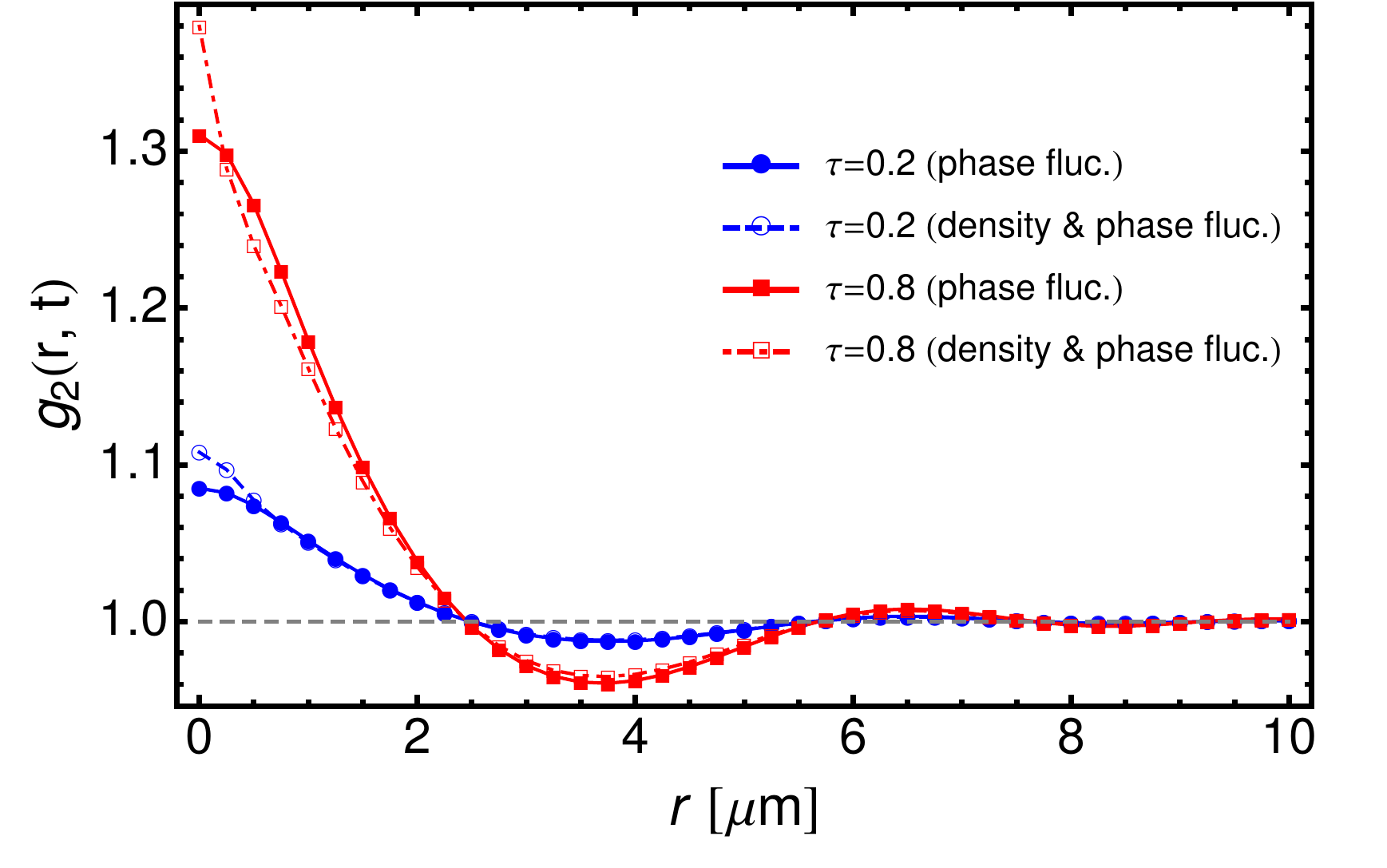}
\caption{(Color online)
 We show $g_2({\bf r},t)$ for the quasi-condensed phase of $^{87}\mathrm{Rb}$ atoms with a density of  $n_0=40\, \mu \mathrm{m}^{-2}$ after two different expansion times $t$. In panel (a) we use $t=1\, {\rm ms}$, and in panel (b) $t=3\, {\rm ms}$.
  The dashed and the dot-dashed lines show the result based on Eq. (\ref{correlation_expression_density_phase}), which takes into account density fluctuations. 
 For comparison, we show the result for the phase fluctuations only, based on Eq. (\ref{correlation_expression}), as solid lines. 
 In particular, the blue line with filled circles and the red line with filled squares correspond to $\tau=0.2$ and $\tau=0.8$, respectively, with phase fluctuations only. 
The blue, dashed line with empty circles and the red, dot-dashed line with empty squares correspond to $\tau=0.2$ and $\tau=0.8$, respectively, with both density and phase fluctuations. 
 For the case of $\tau= 0.2$ we use a de Broglie wavelength of $\lambda_T \approx 0.7\,\mu \mathrm{m}$, and for $\tau=0.8$ we use $\lambda_T \approx 0.4\,\mu \mathrm{m}$.  }
\label{Fig8}
\end{figure}

In Fig. \ref{Fig5}, we compare locations of the spectral peaks obtained from the mean-field results, Eq. (\ref{zero-order_peakpositions}), to the spectral peak locations obtained numerically from Eq. (\ref{spectrum_expression}) for a finite system. We observe that the spectral peak locations are shifted towards lower wavevectors $q$ than the mean-field locations. However, the shift of the spectral peak locations becomes negligible for large peak order numbers $n$.  

  Figure \ref{Fig6} shows the shift of the spectral peak locations on semi-logarithmic scale for different peak order numbers $n$ after various expansion times for the quasi-condensed phase. Here, we show a comparison between the spectral peak shifts obtained numerically from Eq. (\ref{spectrum_expression}) for a finite system and the shifts obtained from the analytic expression, Eq. (\ref{peak_shift}), for a cloud of $^{87} \mathrm{Rb}$ atoms. Both the analytic results and the numerical ones are in good agreement.
The dashed lines in Fig. \ref{Fig6} are the results obtained from the analytic expression, Eq. (\ref{approximated_peak_positions}), that approximates the scaling behavior for the spectral peak shift.

Overall we find a nearly linear behavior due to the mean-field contribution, with a small, negative  $\sim 1/n^{2}$ correction, as can be seen from Eq. (\ref{approximated_peak_positions}), which is due to the fluctuations and the short-range cut-off.

\begin{figure}[] 
\includegraphics[width=0.85\linewidth]{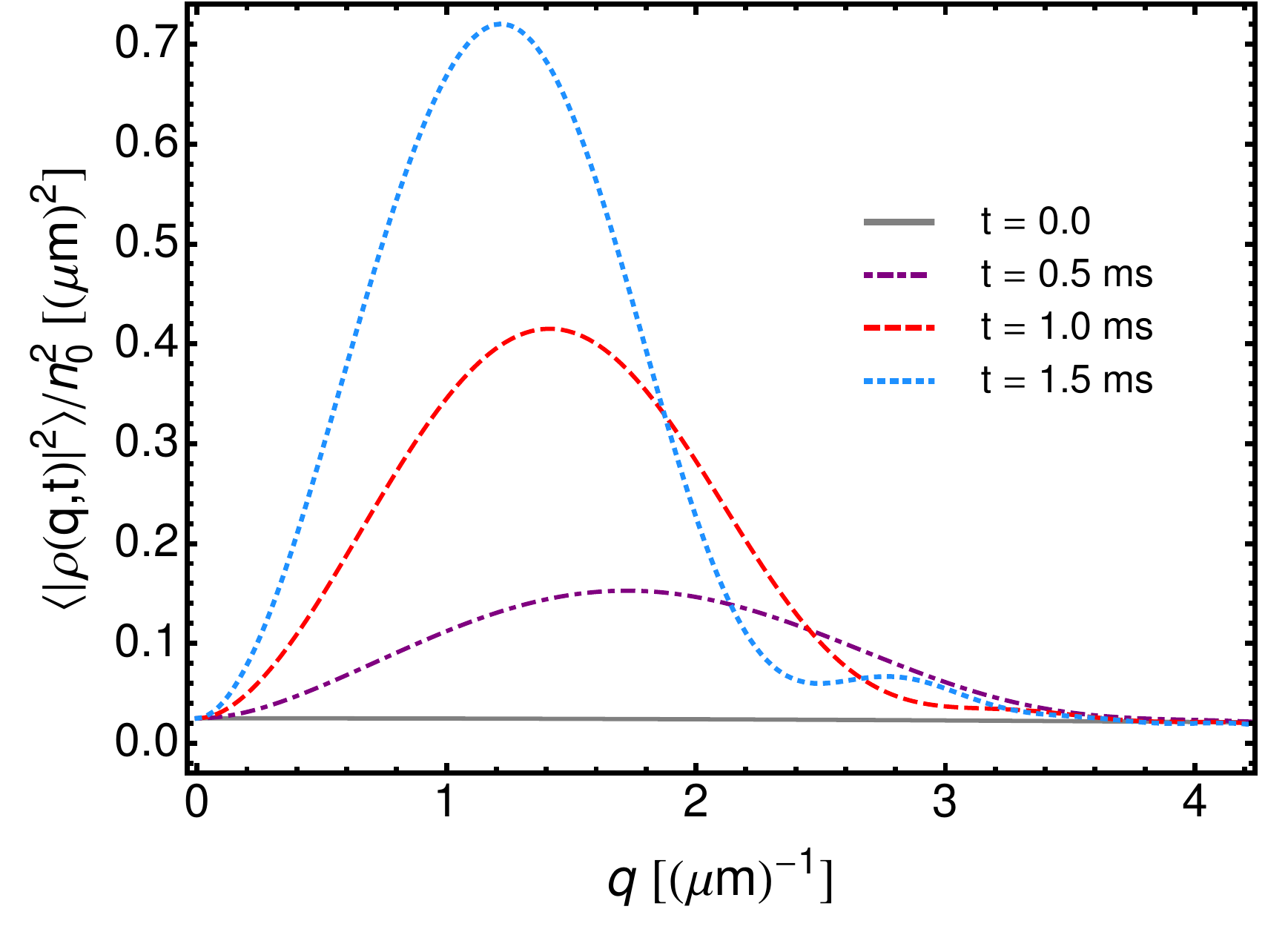}
\caption{(Color online). The spectrum of density-density correlations $\braket{|\rho({\bf q})|^2}/n_0^2$, including  the contribution due to in-situ density fluctuations, for a quasi-condensate of $^{87}\mathrm{Rb}$ atoms with a density  of $n_0=40\, \mu \mathrm{m}^{-2}$ after several expansion times $t$, for $\tau=0.8$. 
The gray, solid line corresponds to the original quasi-condensate ($t=0$); 
the purple, dot-dashed line corresponds to $t=0.5\, {\rm ms}$; 
the red, dashed line corresponds to $t=1\, {\rm ms}$;
the light blue, dotted line corresponds to $t=1.5\, {\rm ms}$.
The thermal de Broglie wavelength is $\lambda_T \approx 0.4\,\mu \mathrm{m}$.  }
\label{Fig9}
\end{figure}
%
%

%
\subsection{Power spectrum above the BKT transition } \label{spectrumexponential_subsection}
 	Now we discuss the spectrum of the density-density correlations of 2D Bose gases above the BKT transition (thermal phase). We calculate the spectrum of the  density-density correlations numerically using Eq. (\ref{spectrum_expression}) for a finite size system. Figure \ref{Fig7} shows the spectrum of the density-density correlations of an expanded cloud of $^{87}$Rb atoms for three different correlation lengths $r_0$ after $0.5\, {\rm ms}$ and $3\, {\rm ms}$ expansion times.  
%

%
%
\section{Influence of density fluctuations on density correlations} \label{density_fluctuation_section}
	So far, we considered only the phase fluctuations of the condensed phase, as expressed by Eq. (\ref{g1_quasicondensed}). In this section, we include the density fluctuations  and analyze their effect on the density-density correlations and the spectrum of these correlations in time-of-flight expansion. 
  We consider small fluctuations of density, and therefore expand $\sqrt{\hat{n}}$ up to second order in $\delta \hat{n}$ as
\begin{equation} \label{density_flucuations_ex}
\sqrt{n_0({\bf r}) +\delta\hat{n}({\bf r})} \approx \sqrt{n_0} \Bigl[1 + \frac{1}{2} \frac{\delta\hat{n}}{n_0} - \frac{1}{8} \frac{\delta\hat{n}^2}{n_0^2} \Bigr]. 
\end{equation}  
  The contribution of small fluctuations of density to the two-point correlation function is \cite{mora_2003}  
\begin{equation} \label{g1_phase_density_fluc_def}
g_1({\bf r}) \cong \mathcal{F}_a({\bf r})  \Bigl[1- \frac{1}{8} \braket{(\Delta \delta\tilde{n})^2}  \Bigr],
\end{equation}  
where $\Delta \delta\tilde{n} \equiv \delta\tilde{n}({\bf 0}) - \delta\tilde{n}({\bf r})$ and $\delta\tilde{n}({\bf r}) = \delta\hat{n}({\bf r})/n_0({\bf r})$. We calculate the expectation value of $(\Delta \delta\tilde{n})^2$  using Bogoliubov theory.
 We find that for temperatures small compared to the chemical potential, $k_BT<\mu$, the density fluctuations are suppressed, and
 the two-point correlation function is principally governed by phase fluctuations.
 This corresponds to the regime $\lambda_T > \xi$,  where $\lambda_T$ is the thermal de Broglie wavelength, $\lambda_T \equiv \sqrt{2\pi \hbar^2/(mk_BT)}$, and $\xi$ is the healing length, related to the chemical potential $\xi= \hbar/\sqrt{m \mu}$. 
   For  temperatures $k_BT>\mu$, i.e., $\lambda_T<\xi$, density fluctuations are of sizeable magnitude.
 They are predominantly due to the thermal single particle excitations of the Bogoliubov spectrum, and are approximately   
\begin{equation}\label{density_fluc_corr}
\braket{\delta\tilde{n}({\bf r}_i) \delta\tilde{n}({\bf r}_j)} \approx \frac{1}{n_0 \lambda_T^2} e^{- \frac{\pi |{\bf r}_{i}-{\bf r}_{j}|^2}{\lambda_T^2}}.
\end{equation} 
%
 With this, Eq. (\ref{g1_phase_density_fluc_def}) becomes    
\begin{equation} 
g_1({\bf r}) \cong \mathcal{F}_a({\bf r}) \Bigl[1 - A \bigl(1 -e^{- \frac{\pi |{\bf r}|^2}{\lambda_T^2}}  \bigr)  \Bigr],
\end{equation}  
where $A = 1/(4n_0 \lambda_T^2)$.

\begin{figure}[] 
\hspace{-2.5 in}(a) \\
\includegraphics[width=0.75\linewidth]{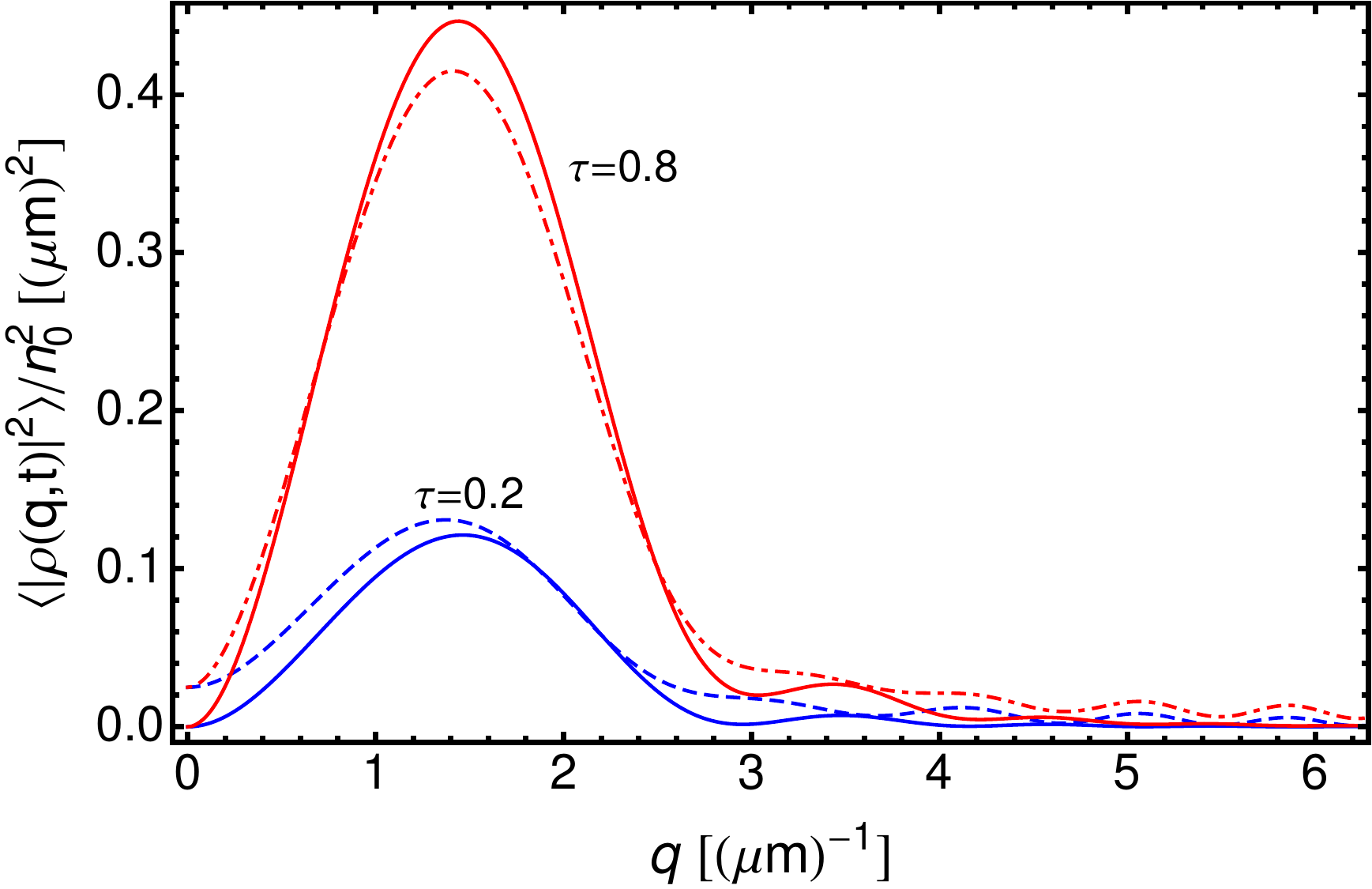} \\
\hspace{-2.5 in}(b)\\
 \includegraphics[width=0.75\linewidth]{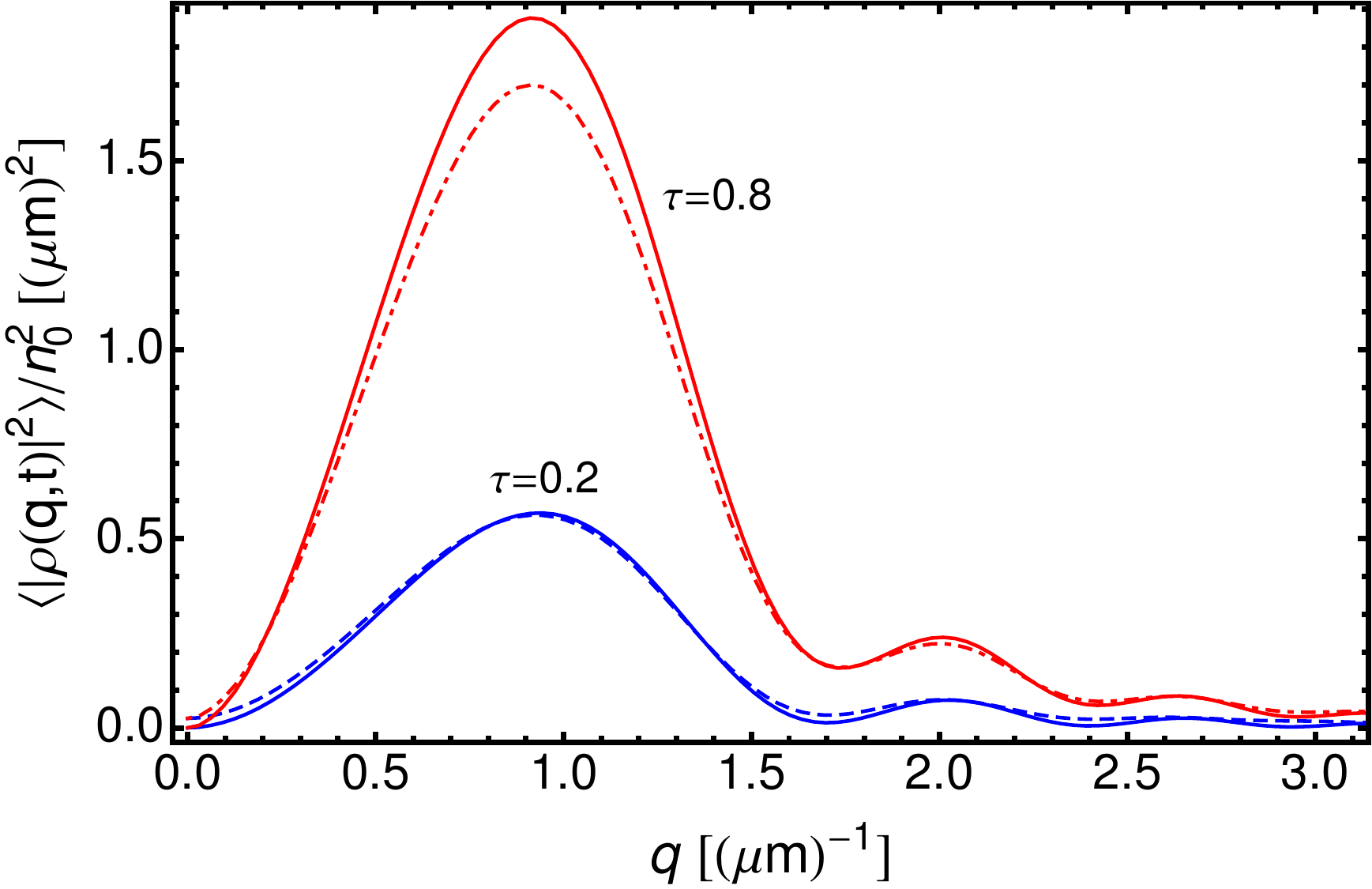}
\caption{(Color online) Power spectrum $\braket{|\rho({\bf q})|^2}/n_0^2$ of a quasi-condensate of $^{87}\mathrm{Rb}$ atoms with density $n_0=40\, \mu \mathrm{m}^{-2}$ after two different expansion times $t$,  
(a) $t=1\, {\rm ms}$, and 
(b) $t=3\, {\rm ms}$.  
The blue, solid line and the red, solid line correspond to  $\tau=0.2$ and $\tau=0.8$, respectively. These show the spectrum for phase fluctuations only, see Eq. (\ref{spectrum_expression}). 
The blue, dashed-line and the red, dot-dashed line correspond to $\tau=0.2$ and $\tau=0.8$, respectively, with both density and phase fluctuations, based on the Fourier transform of Eq. (\ref{correlation_expression_density_phase}). 
For the examples with $\tau=0.2$ we use  $\lambda_T \approx 0.7\,\mu \mathrm{m}$, for $\tau=0.8$ we use $\lambda_T \approx 0.4\,\mu \mathrm{m}$.}
\label{Fig10}
\end{figure}
\begin{figure}[] 
\hspace{-2.5 in}(a) \\
\includegraphics[width=0.75\linewidth]{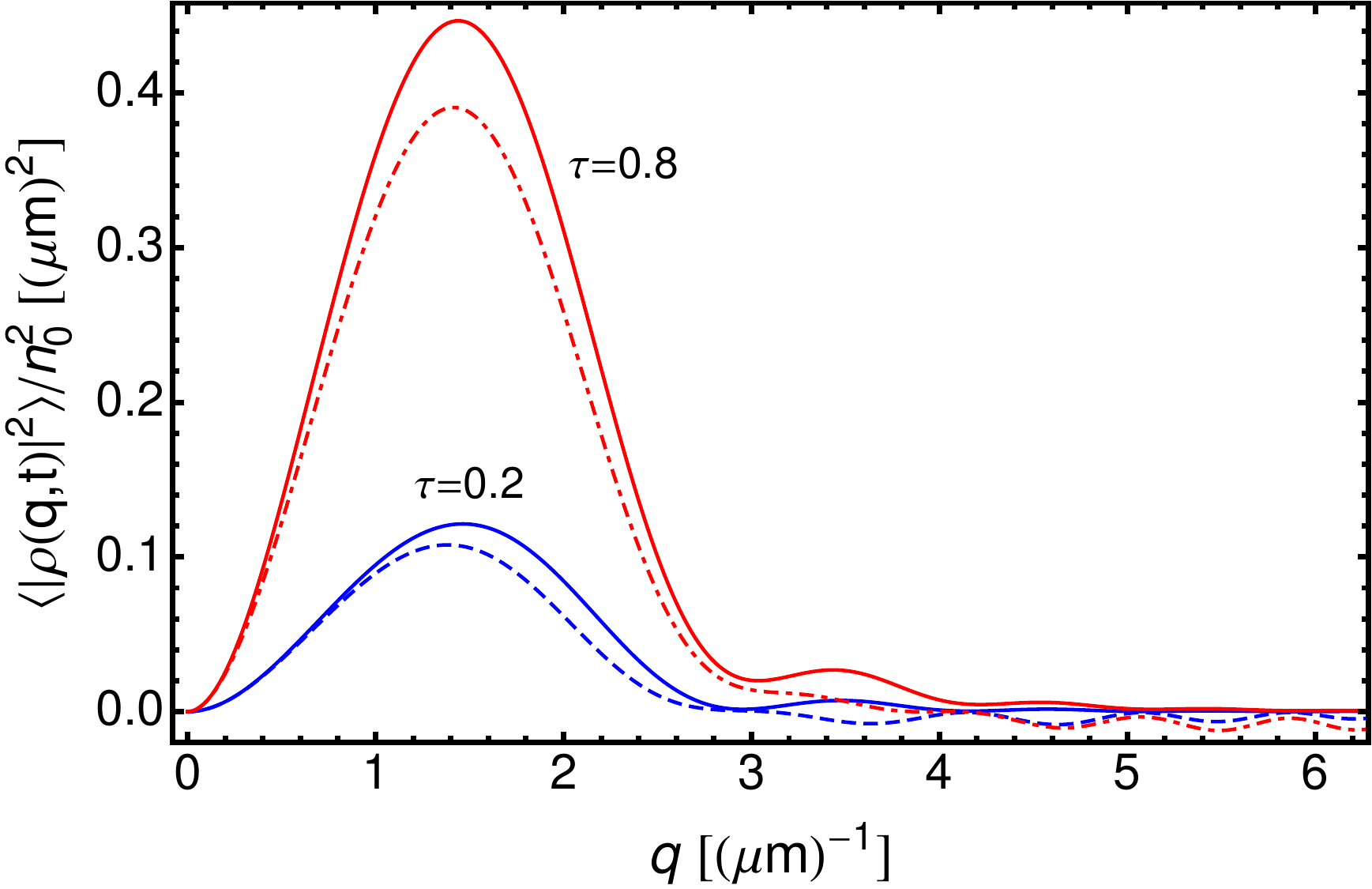} \\
\hspace{-2.5 in}(b)\\
 \includegraphics[width=0.75\linewidth]{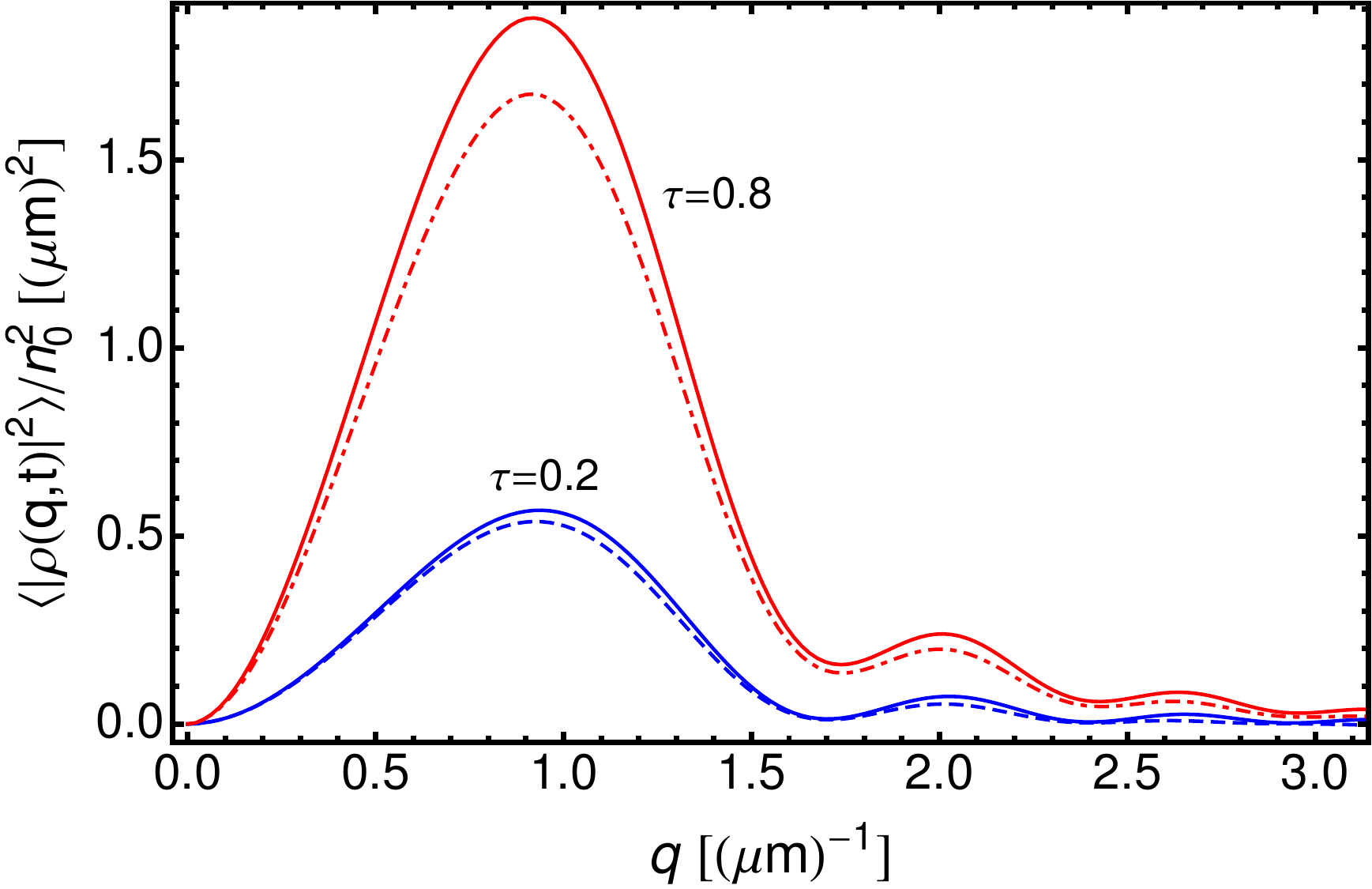}
\caption{(Color online)
 The solid lines are the same data as in Fig. \ref{Fig10}, the power spectrum 
 of a quasi-condensate of $^{87}\mathrm{Rb}$ atoms with density $n_0=40\, \mu \mathrm{m}^{-2}$, with only phase fluctuations taken into account. 
The blue, dashed-line and the red, dot-dashed line also correspond to the same parameters as in Fig. \ref{Fig10}, however, now the contribution due to in-situ density fluctuations has been subtracted, Eq. (\ref{spectrum_background}). 
}
\label{Fig11}
\end{figure}

   In order to calculate the corrections to $g_2({\bf r},t) $ due to density fluctuations, we consider
    Eq. \ref{correlation_expression_general}, and expand the product of four single particle operators to second order in $\delta\tilde{n}$,
    using Eq. (\ref{density_flucuations_ex}). Here we focus on the high-temperature contributions, and ignore further shot noise contributions due to the commutators between $\delta\tilde{n}$ and $\tilde{\theta}$.
 With this, $g_2({\bf r},t)$  is given by 
\begin{align}\label{correlation_expression_density_phase}
&g_2({\bf r},t) = \frac{1}{(2\pi)^2} \int d^2{\bf q} \int d^2{\bf R} \cos {\bf q\cdot r} \cos {\bf q\cdot R} \nonumber\\
&\times \Biggl( \frac{\mathcal{F}_a({\bf q}_t)^2 \mathcal{F}_a({\bf R})^2}{\mathcal{F}_a({\bf R} - {\bf q}_t) \mathcal{F}_a({\bf R}+ {\bf q}_t)}  \Biggr) 
\Biggl( 1 +  \frac{1}{2} \Bigl[ \braket{\delta\tilde{n}({\bf R}) \delta\tilde{n}({\bf 0})} \nonumber \\
&\quad + \braket{\delta\tilde{n}({\bf q}_t) \delta\tilde{n}({\bf 0})} \Bigr]  + \frac{1}{4} \Bigl[ \braket{\delta\tilde{n}({\bf R}-{\bf q}_t) \delta\tilde{n}({\bf 0})} \nonumber \\
&\quad + \braket{\delta\tilde{n}({\bf R}+{\bf q}_t) \delta\tilde{n}({\bf 0})} \Bigr] - \frac{1}{2} \braket{\delta\tilde{n}({\bf 0})^2} \Biggr).
\end{align}
We use this expression to calculate $g_2({\bf r}, t)$ numerically for a finite system, as described above. In Fig. \ref{Fig8}, we compare the two-point density correlation function  for the density and phase fluctuating quasi-condensate to the phase fluctuating quasi-condensate. We observe that the correlation function is enhanced by density fluctuations for distances $r \lesssim \lambda_T$ and it is mainly given by phase fluctuations for distances $r \gtrsim \lambda_T$, with a small reduction of contrast. We also note that this effect of the density fluctuations decreases with increasing expansion time, because of the small length scale associated with the density fluctuations.    

	Next, we calculate the spectrum of the density-density correlations, including the effect of density fluctuations, i.e.,
 we Fourier transform $g_2({\bf r}, t)$, given in Eq. (\ref{correlation_expression_density_phase}).
 In Fig. \ref{Fig9}, we show the spectrum of the density-density correlations as it evolves for short expansion times. Initially, at $t=0$, the power spectrum is just the spectrum of in-situ density fluctuations, given by
\begin{equation} \label{spectrum_background}
\frac{\braket{|\rho({\bf q})|^2}_0 }{n_0^2} = \frac{1}{n_0} e^{-\frac{q^2\lambda_T^2}{4\pi}}. 
\end{equation}	
 As the atomic cloud expands, the phase fluctuations translate into density ripples, and an oscillatory pattern emerges. 
  Two further examples are shown in Fig. \ref{Fig10}.

  Since we want to extract the scaling exponent $\tau$ which only controls the phase fluctuations, we have to distinguish between the features which are due to the density fluctuations and which are due to the phase fluctuations.
  We note that these two contributions occur on very different length scales. 
 This can be easily seen from the single-particle correlation function shown in Eq. (\ref{g1_phase_density_fluc_def}).
While the phase fluctuations lead to a slowly decaying function $\mathcal{F}_a({\bf r})$, the density correlations fall off quickly on a length scale $\lambda_{T}$, over which the function  $\mathcal{F}_a({\bf r})$ is nearly constant. As a result the phase and the density contributions nearly separate into a sum. 
 A similar observation holds for $g_2({\bf r}, t)$. As a result, 
  the spectrum nearly separates into a sum of two terms, one due to the in-situ density fluctuations and the other due to phase fluctuations. 
  
  We therefore propose to subtract the contribution of in-situ density fluctuations from the spectrum, as shown 
   in Fig. \ref{Fig11}. 
 We observe that the resulting spectrum is essentially just that of a purely phase fluctuating condensate, up to 
  wavevectors $q \lesssim 1/\lambda_T$. Especially for $q \rightarrow 0$, the two lines coincide, indicating that only the long-range contributions due to phase fluctuations are present, as desired.
  Alternatively, if the in-situ density correlations are not known independently, the full spectrum can be fitted with a Gaussian distribution, describing the density fluctuation contribution, and a fitting function of the form in Eq. (\ref{spectrum_1order}), to determine the scaling exponent $\tau$.

\section{conclusions} \label{conclusion_section}
	In conclusion, we have analyzed the density-density correlations of an ultra-cold atomic Bose gas
 in two-dimensions, below and above the BKT transition, during time-of-flight. As a typical example, we considered condensates of $^{87}$Rb atoms. 
We first discussed the density-density correlations and the spectrum of these correlations in time-of-flight taking into account only the phase fluctuations, 
 and later included the contribution due to the density fluctuations. We have shown that the quasi-condensed and the thermal phase of the 2D Bose gases can be distinguished from the interference pattern observed for the density-density correlations.      
 Below the BKT transition, this 
 interference pattern  is controlled by the scaling exponent of the quasi-condensed phase. Above the transition, the oscillating pattern is quickly suppressed for expansion distances beyond the correlation length.
 We propose to use the power spectrum of the density ripples to detect the algebraic scaling exponent $\tau$.
We have demonstrated that the spectrum of the density-density correlations is a superposition of the in situ density fluctuations term and the phase fluctuations term. The analytic expression for the power spectrum without the in situ density fluctuations term can be used as a fitting function to experimentally detect the algebraic scaling exponent of the phase fluctuating 2D quasi-condensates, 
 after removing the contribution due to in-situ density fluctuations.

\section*{Acknowledgements}
We gratefully acknowledge valuable discussions with Jean Dalibard, R\'{e}mi Desbuquois, Selim Jochim, Eite Tiesinga, and Yong-il Shin. We acknowledge support from the Deutsche Forschungsgemeinschaft through the Hamburg Centre for Ultrafast Imaging and the SFB 925, and from the Landesexzellenzinitiative Hamburg, which is supported by the Joachim Herz Stiftung.
%


%
\section*{Appendix A} 
\setcounter{equation}{0}
\renewcommand{\theequation}{A{\arabic{equation}}}
In this appendix, we derive the analytic result, Eq. (\ref{spectrum_1order}), for the quasi-condensed phase. The expression for the spectrum of the density-density correlations for the phase fluctuating quasi-condensate in two-dimensions [Eq. (\ref{spectrum_expression})] is given by
\begin{align}\label{expression_spectrum_A}
\frac{\braket{|\rho({\bf q})|^2}}{n_0^2} = \int d^2{\bf r} \cos {\bf q\cdot r} \Biggl( \frac{\mathcal{F}_a({\bf q}_t)^2 \mathcal{F}_a({\bf r})^2}{\mathcal{F}_a({\bf r} - {\bf q}_t) \mathcal{F}_a({\bf r}+ {\bf q}_t)}   -1 \Biggr),
\end{align}
where ${\bf q}_t \equiv \hbar {\bf q}t/m$ and $\mathcal{F}_a({\bf r})$ is the algebraic decay for the quasi-condensate, which is given by
\begin{equation} \label{g1_quasicondensed_A} 
\mathcal{F}_a({\bf r}) \equiv \left(\frac{a^2}{a^2 + |{\bf r}|^2}\right)^{\tau/8}.
\end{equation}
We consider only the $x$-component ($q_x$) of the wavevector ${\bf q}$ and expand the above integral, Eq. (\ref{expression_spectrum_A}), to first order in the exponent $\tau$. We get the following first order term: 
\begin{align} \label{term_1order}
&\frac{\tau}{8} \mathcal{F}_a(q_{tx})^2 \int dx \cos(q_x x) \int dy \ln \Biggl( \Bigl(a^2 + (x-q_{tx})^2  \nonumber\\
& + s(y)^2 \Bigr)  \Bigl(a^2 + (x+q_{tx})^2 + s(y)^2 \Bigr) \Bigl(a^2 + x^2 + s(y)^2 \Bigr)^{-2} \Biggr)  
\end{align}
with $s(y)$ defined as $s(y)=L/\pi \sin(\pi y/L)$. Here, we explicitly kept the expression for a finite size system and used $q_{tx} \equiv \hbar q_xt/m$.  
  After integrating over the $y$-dimension of the system and Fourier transforming along $x$-axis, we get an analytic expression for the spectrum of the density-density correlations, which is given by   
\begin{align}\label{spectrum_1order_A}
\frac{\braket{|\rho({\bf q})|^2}}{n_0^2} &\approx \frac{\pi a \tau K_1(a q)}{q} \left(\frac{a^2}{a^2+\frac{q^2 \hbar ^2 t^2}{m^2}}\right)^{\tau /4} \nonumber \\
& \quad \times \left(1 - \cos \left(\frac{q^2 \hbar t }{m}\right) \right), 
\end{align}  
which is Eq. (\ref{spectrum_1order}).

\section*{Appendix B} 
\setcounter{equation}{0}
\renewcommand{\theequation}{B{\arabic{equation}}}
%
%
In this appendix, we derive the analytic expressions, Eqs. (\ref{peak_shift}) and (\ref{approximated_peak_positions}), from the analytic result for the spectrum of density-density correlations, Eq. (\ref{spectrum_1order_A}). The first-order correction to the location of the spectral peaks is given by
\begin{align} \label{spectral_peak_shift_A}
\Delta q_{n} &= 2 m^2 (a^2 m^2 q_{n0}+ q_{n0}^3 t^2 \hbar ^2) \Bigl[ 2 a^3 m^2 K_2(a q_{n0})  \nonumber\\
& \quad + q_{n0} t^2 \hbar ^2 \bigl( \tau K_1(a q_{n0})+2 a q_n K_2(a q_{n0}) \bigr) \Bigr]  \nonumber\\
& \quad \times \Bigl[ 4 a m^2 K_2(a q_{n0}) (a^2 m^2+q_{n0}^2 t^2 \hbar ^2)  \nonumber\\
& \quad \times \bigl(3 a^2 m^2 +q_{n0}^2 t^2 (\tau +3) \hbar ^2 \bigr)  \nonumber\\
& \quad  + q_{n0} K_1(a q_{n0}) \Bigl(4 a^6 m^6-2 a^2 m^4 t^2 \tau \hbar ^2     \nonumber\\
& \quad  + m^2 q_{n0}^2 t^4 \hbar ^4 \bigl(\tau  (\tau +2)-12 a^2 q_{n0}^2 \bigr)  \nonumber\\
& \quad  - 8q_{n0}^6 t^6 \hbar ^6 \Bigr) \Bigr]^{-1},
\end{align}
where $K_1$ and $K_2$ are the Bessel functions of second kind.

  After rearranging the above equation and neglecting the second order term in exponent $\tau$, we arrive at
\begin{align} \label{spectral_peak_shift_rewrite_A}
\Delta q_{n} & \approx q_{n0} L_n^2 \Bigl(q_{n0} t^2\tau \hbar^2 K_1(aq_{n0}) + 2a L_n^2 m^2 K_2(aq_{n0}) \Bigr) \nonumber\\
& \quad \times \Bigl[ q_{n0} \Bigl(2a^2 m^2(a^4 - q_{nt}^4) + t^2  \bigl(-2a^2 \tau   \nonumber \\
&\quad +L_n^2(\tau- 4q_{n0}^2 q_{nt}^2 ) \bigr) \hbar^2 \Bigr) K_1(aq_{n0}) + 2 a m^2 L_n^2         \nonumber \\
& \quad  \times \bigl(3L_n^2 +q_{nt}^2\tau \bigr) K_2(aq_{n0}) \Bigr]^{-1},
\end{align}
which is Eq. (\ref{peak_shift}). Here $L_n^2(t) = a^2 +q_{nt}^2$, and $q_{nt} \equiv \frac{q_{n0}\hbar t}{m}$. 
We now consider two cases either $a q_{n0} \ll 1$ or $a q_{n0} \gg 1$. When $a q_{n0} \ll 1$, the Bessel functions in Eq. (\ref{spectral_peak_shift_rewrite_A}) can be replaced as $K_1(X)\sim 1/X$, and $K_2(X)\sim 2/X^2$. The limit $a q_{n0} \ll 1$, i.e.,  $t \gg \frac{m a^2}{\hbar} (2n-1)\pi$ translates $L_n(t) \sim q_{nt}$. And we get
\begin{align}\label{approximated_peak_positions_A}
\Delta q_{n} & \sim - \frac{q_{n0}(4 + \tau)}{ 2 q_{n0}^2 (a^2 + 2q_{nt}^2)-5\tau -12} \nonumber \\
& =  - \frac{q_{n0}(4 + \tau)}{ 4(2n-1)^2 \pi^2 + \frac{2ma^2}{\hbar t}(2n-1)\pi -5\tau -12}, 
\end{align}
which is Eq. (\ref{approximated_peak_positions}).
  When $a q_{n0} \gg 1$, the Bessel functions $K_1(X)$ and $K_2(X)$ can be replaced by their asymptotic values as $K_1(X) = K_2(X) \sim \sqrt{\frac{\pi}{2 X}} \exp(-X)$. Since, $a q_{n0} \gg 1$, i.e., $t \ll \frac{m a^2}{\hbar} (2n-1)\pi$ and $L^2_n(t)=a^2 + \frac{\hbar t}{m}(2n-1)\pi$, these lead to two limiting cases: either $t \ll \frac{m a^2}{\hbar} \frac{1}{(2n-1)\pi}$ or $t \gg \frac{m a^2}{\hbar} \frac{1}{(2n-1)\pi}$. If $t \ll \frac{m a^2}{\hbar} \frac{1}{(2n-1)\pi}$ that means $L_n(t) \sim a$. We obtain 
\begin{align}
\Delta q_{n} & \sim - q_{n0}(2a^3 m^2 +q_{n0} t^2 \tau\hbar^2) \nonumber \\
& \times \Bigl[ q_{n0}t^2(4 q_{n0}^2 q_{nt}^2+\tau)\hbar^2 \nonumber \\
& - 2m^2(3a^3 +a^4q_{n0}-q_{n0}q_{nt}^4+ aq_{nt}^2 \tau) \Bigr]^{-1}.
\end{align}
The condition $t \gg \frac{m a^2}{\hbar} \frac{1}{(2n-1)\pi}$ translates into $L_n(t) \sim q_{nt} $, And we get
\begin{equation}
\Delta q_{n} \sim - \frac{(2 a q_{n0}+ \tau)}{2a(a q_{n0}-\tau -3) + 4q_{n0} q_{nt}^2}.
\end{equation}

\begin{thebibliography}{99}
\bibitem{distancecutoff_a} For a Bose gas in finite two-dimensional systems the short distance cutoff $a$ is defined as $a= \lambda^2_T/(2\pi \xi_h)$, where $\lambda_T$ is the thermal de Broglie wavelength, and $\xi_h$ is the healing length.
\end{thebibliography}
\end{document}